\documentclass[fleqn,usenatbib]{mnras}

\usepackage{url}

\usepackage[T1]{fontenc}
\usepackage{ae,aecompl}
\usepackage[draft]{cleveref}
\usepackage{comment}
\usepackage{pdflscape}

\usepackage{array}

\usepackage{graphicx}	
\usepackage{amsmath}	
\usepackage{amssymb}	
\usepackage{subfig}
\usepackage{caption}

\usepackage{floatrow}
\usepackage{hvfloat}

\usepackage{color}

\usepackage{soul}

\title[TeV gamma-ray discovery of 2WHSP~J073326.7+515354]{Testing emission models on the  extreme blazar 2WHSP~J073326.7+515354 detected at very high energies with the MAGIC telescopes}

\author[The MAGIC Collaboration]{
\parbox{0.99\textwidth}{
\normalsize 
MAGIC Collaboration:
V.~A.~Acciari$^{1}$,
S.~Ansoldi$^{2,23}$,
L.~A.~Antonelli$^{3}$,
A.~Arbet Engels$^{4}$,
D.~Baack$^{5}$,
A.~Babi\'c$^{6}$,
B.~Banerjee$^{7}$,
U.~Barres de Almeida$^{8}$,
J.~A.~Barrio$^{9}$,
J.~Becerra Gonz\'alez$^{1}$
% ...
}\newauthor\normalsize{Affiliations can be found at the end of the article.}
}

\pubyear{2019}

\begin{document}
\label{firstpage}
\pagerange{\pageref{firstpage}--\pageref{lastpage}}
\maketitle

\begin{abstract}
Extreme high-energy peaked BL~Lac objects (EHBLs) are an emerging class of blazars. Their typical two-hump structured spectral energy distribution (SED) {peaks at higher energies} with respect to conventional blazars. Multi-wavelength (MWL) observations constrain their synchrotron peak in the medium to hard X-ray band. Their gamma-ray SED peaks above the GeV band, and in some objects it extends up to several TeV. 
Up to now, only a few EHBLs have been detected in the TeV gamma-ray range. In this paper, 
we report the detection of the EHBL 
2WHSP~J073326.7+515354,
{observed and detected  during 2018} 
in TeV gamma rays with the MAGIC telescopes. The broad-band SED is studied within a MWL context, including an analysis of the \emph{Fermi}-LAT data over ten years of observation and with simultaneous {\it Swift}-XRT, {\it Swift}-UVOT, and {KVA data}.
Our analysis results in a set of spectral parameters that confirms the classification of the source as an EHBL. In order to investigate the physical nature of this extreme emission, different theoretical frameworks were tested to model the broad-band SED. The hard TeV spectrum of 2WHSP~J073326.7+515354 sets the SED far from the energy equipartition regime in the standard one-zone leptonic scenario of blazar emission. Conversely, more complex models of the jet, represented by either a two-zone spine-layer model or a hadronic emission model, better represent the broad-band SED.
\end{abstract}

 
\let\footnote\relax\footnotetext{L. Foffano (luca.foffano@phd.unipd.it), J. Becerra Gonz\'alez (jbecerragonzalez@gmail.com), and M. Cerruti (matteo.cerruti@icc.ub.edu)}

\section{Introduction}
Blazars are active galactic nuclei (AGN) with relativistic jets closely aligned with the 
line of sight of the observer. Their spectral energy distributions (SEDs) generally {consist} of two main non-thermal components. Typically, the first component
is ascribed to synchrotron radiation emitted by relativistic electrons moving within the jet.
{Different} scenarios have been proposed to explain the nature of the second hump peaking at higher energies. 
The standard leptonic scenario suggests that this second hump is produced by inverse Compton (IC) scattering of low-energy photons \citep{IC}. In the Synchrotron-Self-Compton (SSC) model \cite[e.g.,][]{Maraschi:SSC,Tavecchio:SSC}, 
this photon field may be composed by the synchrotron emission responsible for the first SED hump. Additionally, this high-energy hump might be associated with external photon fields that are up-scattered by IC scattering in the External Compton scenario \citep{EC}.
\begin{table*}
\renewcommand{\arraystretch}{1.37} 
\scalebox{0.83}{
\hspace{-10pt}
\centering
\begin{tabular}{lccccccccc}
	\hline
 Source name  & R.A.  & $\delta$& Redshift  & Obs. time     & Significance & Integral flux > 200 GeV &   $\Gamma_{\text{obs}}$  & $\Gamma_{\text{intr}}$ \\
      &  deg & deg  &       &         &  & $10^{-13}$ \text{ph}\;$ \text{cm}^{-2}\;\text{s}^{-1}$& & \\
	\hline
2WHSP~J073326.7+515354    & 113.36125 &  	51.89889 &  0.065   & 23.38 h      & 6.76 $\sigma$ & $22.5 \pm 0.60$ &  $2.41\pm0.17$  & $1.99\pm0.16$\\
 \hline
\end{tabular} 
}
\caption{Summary of the observational results obtained with the MAGIC telescopes. We report here the source name, its coordinates, and the first estimation of redshift reported by \citet{PGC-redshift-atel}. The information related to MAGIC observations includes the observation time, the resulting significance of the detection, and the integral flux above 200 GeV. Finally, the observed spectral index $\Gamma_{\text{obs}}$ as measured by MAGIC  is reported together with the intrinsic one $\Gamma_{\text{intr}}$, deabsorbed with the EBL model by \citet{Dominguez:2010bv}.  }
\label{tab:srctable}
\end{table*}

Relativistic protons might also be accelerated in the blazar jet. When sufficiently high energies are reached to allow photo-pion production, electromagnetic cascades will develop and contribute to the emission of the high-energy hump, in addition to proton, muon, and pion synchrotron radiation \citep{Mannheim:1993, Boettcher2010}. Moreover, in the so-called hadronic cascade scenario, {ultra-high energy cosmic rays (UHECRs) might interact in the intergalactic space through photo-hadronic reactions  and produce photons that contribute to the high-energy hump} \cite[e.g.,][]{essey2010,  Murase,tavecchio-cascade-scenario}. 
{Finally, this second hump may be also produced by a combination of leptonic and hadronic processes.}

Blazars are historically subdivided into two main \mbox{categories}. The objects that show broad emission 
lines in { their optical} spectrum are classified as Flat Spectrum Radio Quasars (FSRQs). {When these lines have an equivalent width of less than $5$\,\AA, blazars are defined as BL~Lac objects.}
It has been suggested that blazars follow the so-called ``blazar sequence'' \citep{fossati98}, based on the anti-correlation between the bolometric luminosity and the peak energy of their SED humps \citep{Ghisellini:1998,blazarsequence08,blazarsequence17}. { Conversely, some authors argue that the blazar sequence might be due to selection effects \cite[see e.g.,][]{2005MNRAS.356..225A,2012MNRAS.420.2899G}}. The FSRQs, whose synchrotron peak
is located at low frequencies, are the ``redder'' blazars. The BL~Lac objects populate the sequence at higher frequencies. {Blazars} are further divided in sub-classes depending on the frequency of the synchrotron peak $\nu_{\text{peak}}^{\text{sync}}$: they are classified as low-peaked objects (LBL, with $\nu_{\text{peak}}^{\text{sync}}<10^{14}$ Hz), intermediate-peaked objects (IBL, with $\nu_{\text{peak}}^{\text{sync}}$ between $10^{14}$ and $10^{15}$ Hz), and high-peaked objects (HBL, $\nu_{\text{peak}}^{\text{sync}}$ between $10^{15}$ and $10^{17}$ Hz), {according to \cite{2010ApJ...716...30A}.} 

\citet{2001AA...371..512C} proposed a new class of BL~Lac  objects  with extreme spectral properties and located at the very edge of the blazar sequence, named extreme high-frequency peaked blazars (EHBLs). 
In this work, we will use the definition of EHBL based on the synchrotron peak position $\nu_{\text{peak}}^{\text{sync}}$ located above~$10^{17}$~Hz.

The archetypal EHBL is 1ES~0229+200.

Its archival SED {has been observed in detail} by several multi-wavelength (MWL) observational campaigns during the last years, and shows the {key features} of this class of objects.   
{In fact, in the EHBLs} the synchrotron hump is shifted towards high energies with respect to conventional blazars, making the thermal optical radiation of the host galaxy visible for low-redshift objects.

The synchrotron peak located in the medium-to-hard X-ray~band pushes the second SED peak to the very-high-energy \mbox{gamma-ray} band (VHE, energies above 100 GeV).  For this reason, EHBLs are generally supposed to be faint in high-energy (HE, energies between 100 MeV and 100 GeV) gamma rays \citep{Tavecchio:2009zb}. 
The {intrinsic spectrum} of 1ES~0229+200 at VHE is the hardest ever measured \cite[e.g.,][]{2007A&A...475L...9A,veritas0229}.  
Similar hard spectra have been reported for few other sources, like for example \mbox{1ES~0347-121} \citep{0347discovery}, RGB~J0710+591 \citep{0710discovery}, and 1ES~1101-232 \citep{1101discovery}. 
{In these objects,  the peak of the second hump  extends beyond several TeV}, and for this reason they have {also} been called ``hard-TeV blazars'' by \citet{Nustar_EHBLs}. 

This feature makes their SEDs {challenging} for the standard  one-zone leptonic SSC model. In that scenario, the model would suggest rather soft SSC spectra at TeV energies due to the decreasing scattering cross section with energy in the Klein-Nishina regime  \citep{tavecchio2009}. In order to explain such a shift in the SED peaks, the minimum Lorentz factor of the electron energy distribution $\gamma_\text{min}$ {has to be very high} and the magnetic field intensity $B$ is required to be very low with respect to the standard values inferred in classical TeV BL~Lac objects \citep{Tavecchio:2009zb, Lefa2011}.

To explain the hard-TeV spectra in EHBLs, different alternative models have been proposed. \citet{tev_maxwellian_distribution} and \citet{Lefa2011}, for example, adopt extremely hard Maxwellian particle distributions, while \citet{Katarzynski2006} and \citet{tavecchio2009} use a low-energy cut-off of the electron distribution at VHE.
In the case of 1ES~0229+200,
the intergalactic cascades scenario  \citep{Murase} {was successfully applied to explain} the hard TeV spectrum.
Finally, thanks to the evidence of scarce and low-amplitude flux variability and their hard TeV gamma-ray spectra, EHBLs turn out to be interesting candidates for hadronic and lepto-hadronic emission models, that can well reproduce their observed SEDs \cite[e.g.,][]{Murase, cerruti2015}.

{ The hard VHE gamma-ray spectrum of EHBLs extending up to several TeV -- as that observed in sources like 1ES~0229+200 -- is also an important probe for testing models of the extragalactic background light \citep[EBL, see e.g.,][]{Hauser:2001} and of the intergalactic magnetic fields \citep[IGMF, e.g.,][]{vovk12}.    }

MWL observations have revealed that other EHBL objects have high synchrotron peak frequencies similar to 1ES 0229+200, but much softer TeV spectra with an IC hump clearly peaking in the GeV to TeV band \cite[e.g.,][]{Costamante:2001pu,Nustar_EHBLs}. Additionally, some very bright HBL sources (like, for example, Mrk~501) have shown EHBL-like behavior during some flaring episodes \citep{harder-when-brighter, mkr501-flare2012}. Hence, the EHBL class might be a complex population of sources, characterized by different spectral properties \citep[]{foffano2018}, 
or even associated to high-activity states of some blazars.

Hard-TeV blazars are the EHBL sources with the highest IC peak frequency, 
and the difficulties in modeling their SEDs are generally related to this extreme spectral property. However, EHBLs with a more moderate IC peak located below a few TeV might be good candidates for { testing} theoretical models. {The successful application of theoretical models to different EHBLs might help in understanding why the high synchrotron peak is not always correlated with a hard VHE spectrum, and might help to unveil} the origin of the extreme particle acceleration mechanism of this class.
The differences in the spectral properties  we find in the EHBL category and the  low number of known objects of this class motivate their monitoring and the search for new candidates.

An accurate description of the broad-band spectrum is essential to understand the origin of the extreme SED properties of EHBLs, especially in the gamma-ray band. 
For example, dedicated studies have been recently carried out in the HE gamma-ray band performing detailed analyses of faint \mbox{\emph{Fermi}-LAT} sources \citep{Arsioli:2018yeq}.
In this framework, the TeV gamma-ray band plays a key role in the EHBLs characterization. However, up to now only a few such sources have been observed and characterized in the VHE \mbox{gamma-ray} regime. New TeV observations of EHBL objects are needed in order to increase the EHBL population and possibly disclose the physical interpretation of such extreme spectral properties.

In this paper, we provide a set of new VHE gamma-ray observations on an EHBL named 2WHSP~J073326.7+515354. This blazar, also named PGC~2402248, has been selected from the 2WHSP catalogue \citep{2whsp} on the basis of its high synchrotron peak frequency equal to \mbox{$\nu_{\text{peak,2WHSP}}^{\text{sync}}=10^{17.9}$~Hz}. It is associated with the \emph{Fermi}-LAT source 3FGL~J0733.5+5153 in the 3FGL catalogue \citep{fermi3fgl} as active galaxy of uncertain type, and reported in the 3FHL catalogue \citep{fermi3FHL} as associated with the source 3FHL~J0733.4+51523 with a spectral index of  \mbox{$\Gamma_{HE} = 1.34 \pm 0.43$}.  
Additionally, the source 2WHSP~J073326.7+515354 is compatible with the position (at 2.4 arcmin) of the source SWIFT~J0733.9+5156 (position uncertainty 5.67 arcmin), that is reported also in the  \textit{Swift}-BAT 105-months catalogue \citep{BATcatalog105}. {In this catalogue, the reported flux of the source is $8.17_{6.00}^{10.44} \times 10^{-12}$  \text{erg}\;$ \text{cm}^{-2}\;\text{s}^{-1}$ with a spectral index of~$2.32_{1.61}^{3.25}$}.

The MAGIC observations led to the first detection of this source in TeV gamma rays on 2018 April 19 \citep{PGC-discovery-atel}. During the MAGIC pointings, simultaneous observations were performed by the { KVA, \emph{Swift}-UVOT/XRT}, and \emph{Fermi}-LAT telescopes. Additionally, optical data were collected with the Gran Telescopio Canarias (GTC) in order to estimate the redshift of the source that was previously unknown.
The new measurement of the redshift of 2WHSP~J073326.7+515354 was reported as $z=0.065$ \citep{PGC-redshift-atel}. This value is particularly important for the estimation of the intrinsic gamma-ray spectrum of the source, and consequently for testing the theoretical emission models of the broad-band SED.

The structure of this paper is the following: in \Cref{sec:observations}, we describe the observations and results from the MAGIC observations. In addition, simultaneous observations performed by KVA, \emph{Swift}, and the {long-integration} analysis of the \emph{Fermi}-LAT telescope data are presented. In \Cref{sec:variability} the variability at different frequencies is discussed. In \Cref{sec:sed}, we report the collected broad-band SED and a discussion about the observational properties of the source. In \Cref{sec:modeling}, we provide a discussion on the modeling of the SED, performed by means of leptonic and hadronic models. Finally, we report in \Cref{sec:concl} the conclusions of this work and future prospects.
{We adopt $\mathrm{H}_0 = 67 \,\mathrm{km}\,\mathrm{s}^{-1}\,\mathrm{Mpc}^{-1}, \Omega_{\Lambda} = 0.7, \Omega_{\mathrm{M}} = 0.3$ \citep{plank2018}.}

\section{Observations and results}
\label{sec:observations}

2WHSP~J073326.7+515354 was observed in the VHE gamma-ray band with the MAGIC telescopes, in the optical and UV bands with the KVA telescope and {\it Swift}-UVOT and in the X-ray band with {\it Swift}-XRT. Additionally, { an analysis of the sample} collected by {\it Fermi}-LAT during more than ten years of operation was performed.

\subsection{The MAGIC telescopes}
\label{sec:magic}

MAGIC \citep{magicperf_1:2015} is  a system of two Imaging Air-shower Cherenkov Telescopes (IACTs)
{designed to indirectly detect gamma rays through the Cherenkov emission of the charged component of the extensive air shower they generate interacting with Earth's atmosphere.}
The two telescopes are located on the Canary island of La Palma, at 2200\,m altitude. Their large reflective surface of 17\,m diameter each allows the MAGIC telescopes to reach, under good observational conditions, an energy threshold as low as 50 GeV when operated in standard trigger mode.
The integral sensitivity for point-like sources above 220\,GeV, assuming a Crab Nebula-like spectrum, is ($0.66\pm0.03$)\% of the Crab Nebula flux in 50\,h of observations. At those energies the angular resolution is $0.07$ degree, while the energy resolution reaches 16\%. The performance of the instrument and the details on the data analysis procedure are fully described in \citet{MAGICsens} and references therein.


\subsubsection{Observations}

MAGIC observed the source 2WHSP~J073326.7+515354 for a total of 23.4 h in 2018 within an observational program aimed at searching for new EHBLs in the TeV gamma-ray band.
The observations were performed during 25 nights from 2018 January 23 to April 19 (MJD 58141-58227), with zenith angle range between 23$^\circ$ and 40$^\circ$ and good data quality.
The data have been analysed using
the MAGIC Analysis and Reconstruction Software \cite[MARS,][]{moralejo2009,MAGICsens}.

\begin{figure}
\centering
\subfloat[$\theta^2$ plot.]{\includegraphics[width=\columnwidth]{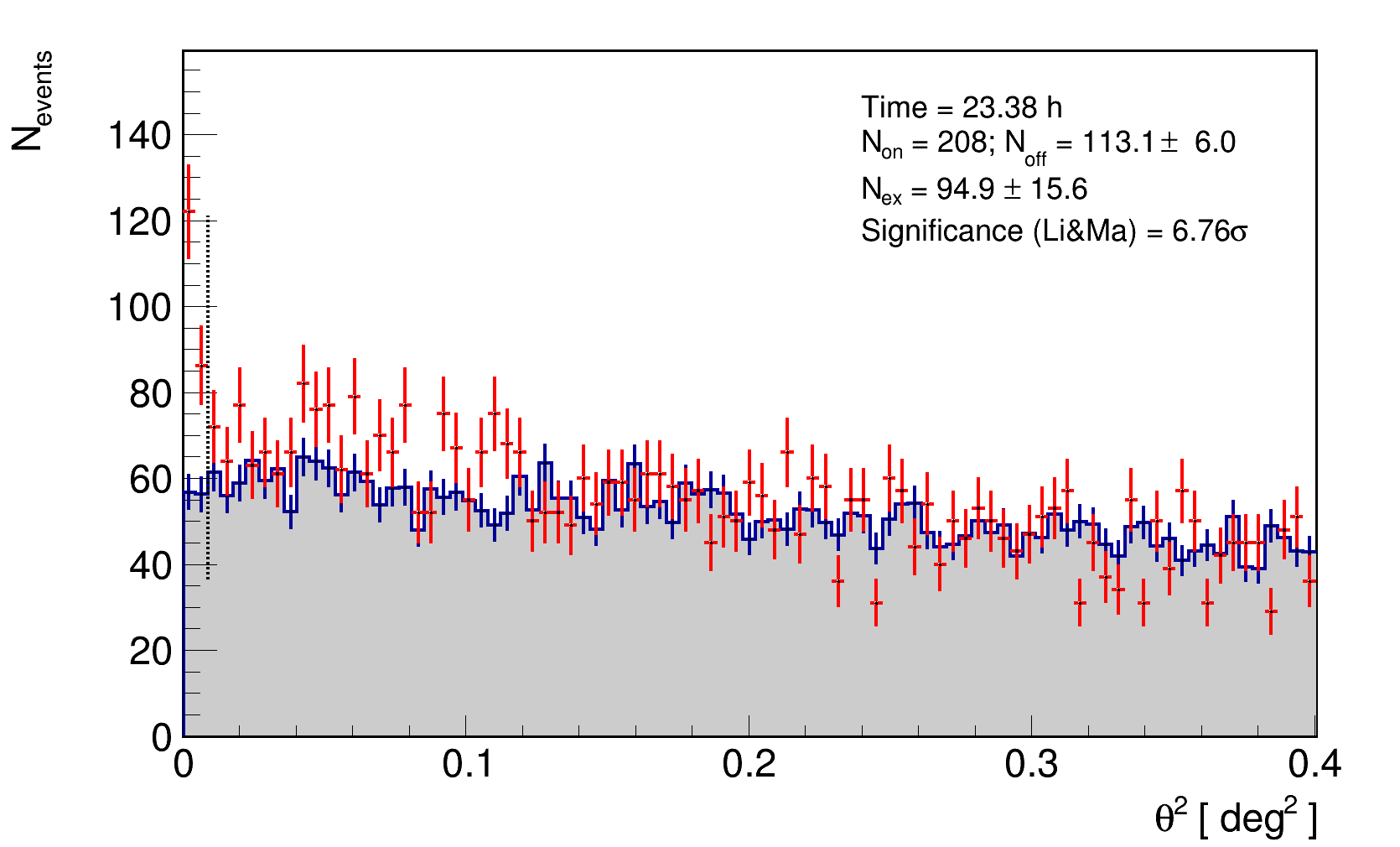}\label{fig:odieFR}}\\
\subfloat[Residuals plot.]{\includegraphics[scale=0.12]{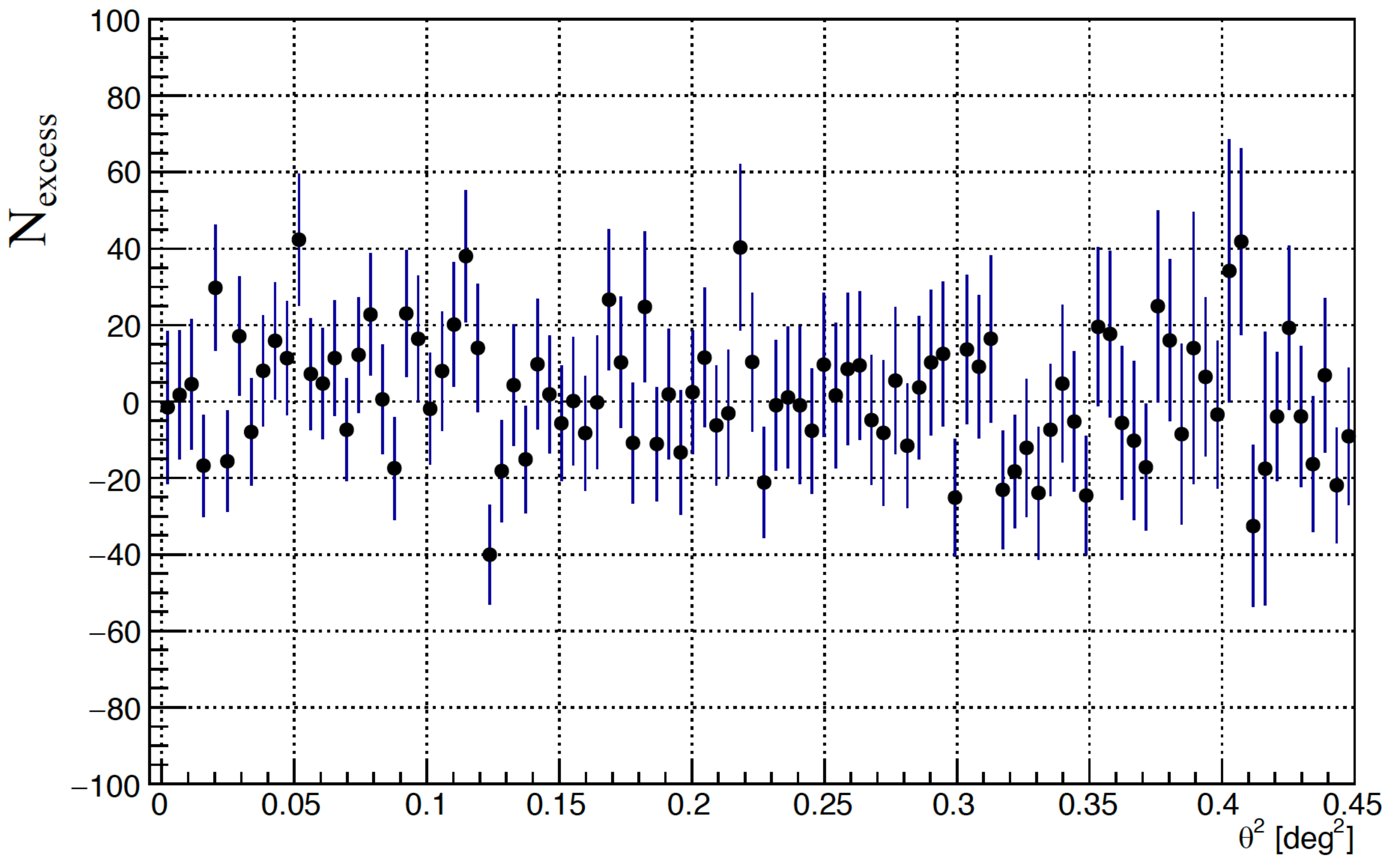}\label{fig:odieFRresiduals}}
\caption{In (a) the $\theta^2$ distribution from the direction of 2WHSP~J073326.7+515354 as observed by the MAGIC telescopes. The gamma-ray like events are represented by the red markers, while the background is denoted by the shadowed grey area. The vertical dashed line indicates the defined signal region to which the significance of the detection is calculated. {In (b) the residuals of the observed data with respect to the fit with the reference PSF of the instrument.}}
\label{fig:odie}
\end{figure}

\vspace{10pt}

\subsubsection{Signal search}\label{subsec:MAGICsignal}

The emission from a source in VHE gamma rays can be evaluated by means of the so-called $\theta^2$ plot. The $\theta^2$ parameter is defined as the squared angular distance between the reconstructed incoming direction of the gamma-ray event and the nominal position of the source in camera coordinates. The typical signature of { VHE point-like sources,} after the application of energy-dependent background suppression cuts, is an excess at low $\theta^2$ values. In general, a source is considered detected in the VHE gamma-ray range, if the significance of the excess of  gamma-like events over background events exceeds~5\,$\sigma$. The significance of the gamma-ray signal is estimated with formula n.17 of \citet{LiMa83}. 

The $\theta^2$ plot for 2WHSP~J073326.7+515354 is shown in \Cref{fig:odieFR}. An excess of $95 \pm 16$ events in the standard fiducial signal region with $\theta^2 < 0.009$~$\mbox{deg}^2$ is found, corresponding to a significance of $6.76\,\sigma$. The $\theta^2$ distribution shows a fluctuation of the gamma-like events with respect to the background events in the region from 0.04 to 0.12 $\mbox{deg}^2$. In order to investigate whether this fluctuation is significant, we compare the $\theta^2$ distribution for 2WHSP~J073326.7+515354 with regards to the reference Point Spread Function (PSF) obtained from a Crab Nebula data sample observed contemporaneously to 2WHSP~J073326.7+515354. The PSF was also rescaled to the 2WHSP~J073326.7+515354 spectrum and zenith distribution. Following \citet{paolodavela-psf}, the PSF and the $\theta^2$-plot of 2WHSP~J073326.7+515354 were fitted with the King function, and a comparison among the parameters was performed. {The fit has been performed up to $\theta^2=0.45$ deg$^2$}. The PSF computed for 2WHSP~J073326.7+515354 is consistent ($\chi^2/\text{DOF} = 89/98$) with the reference PSF of the instrument. {In \Cref{fig:odieFRresiduals} the residuals plot of the fit is shown. This check confirms that the possible mismatch with the background in the region from 0.04 to 0.12 $\mbox{deg}^2$ is not statistically significant, and represent a casual {fluctuation} of the background.}\\

\begin{figure}
\centering
\includegraphics[width=\columnwidth]{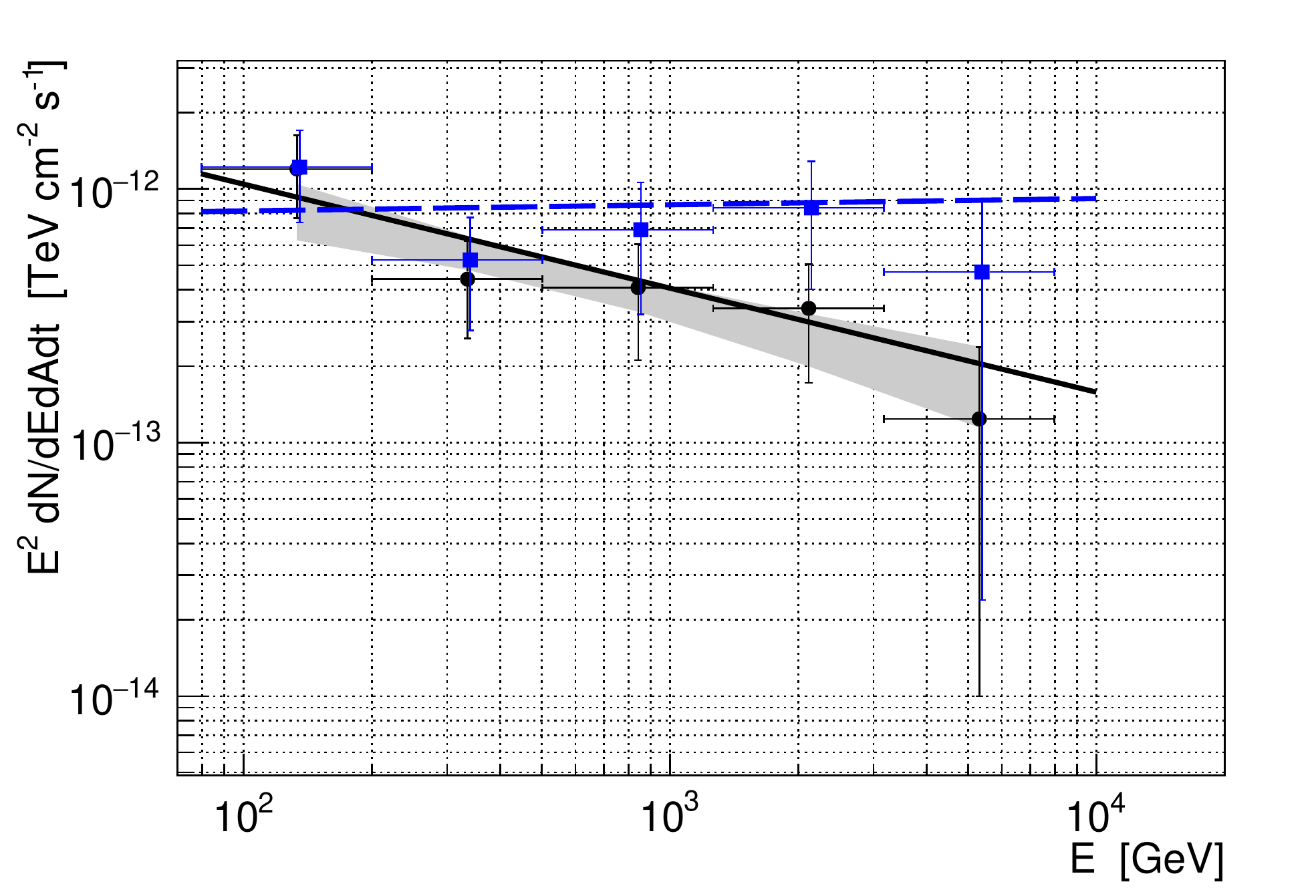}
\caption{VHE SED from 2WHSP~J073326.7+515354. The black markers and line represent the observed SED spectral points and fit. The intrinsic SED after correcting for the EBL absorption assuming the model from \citet{Dominguez:2010bv} is represented in blue. {The shaded area represents the uncertainty obtained from a forward folding method \citep{Mizobuchi:2005qa}. }}
\label{fig:PGC-SED}
\end{figure}

\vspace{10pt}

\subsubsection{Spectrum}  \label{subsec:MAGICspectrum}

The spectrum of the source 2WHSP~J073326.7+515354 observed with the MAGIC telescopes, {reported in \Cref{fig:PGC-SED}}, was reconstructed between 0.1 and 8 TeV using the Tikhonov unfolding method \citep{Albert:2007ss} {in order to include migrations between true and reconstructed energy}. It can be described by a simple power-law model ($\chi^2/\mathrm {DOF}=2.4/3$):
\begin{equation*}
\hspace{70pt} \mathrm{\frac{\mbox{d}N}{\mbox{d}E} = f_{0} \left(\frac{E}{\mathrm{200~GeV}}\right)^{-\Gamma}}\;\;,
\end{equation*}
where the observed photon index is $\mathrm{\Gamma_{\text{obs}}  = 2.41 \pm 0.17_{\mathrm{stat}}}$,
and the corresponding normalization constant $\mathrm{f_{0,\text{obs}}} = (1.95 \pm 0.10_{\mathrm{stat}})
\times 10^{-11}$ \text{ph}\;$ \text{cm}^{-2}\;\text{s}^{-1}\;\text{TeV}^{-1}$  at the
energy of 200\;GeV. {A detailed discussion on the systematic uncertainties can be found in \cite{MAGICsens}.}

The intrinsic spectrum, after correcting for the absorption due to the interaction with the EBL according to the model by \citet{Dominguez:2010bv}, can be fitted with a power-law function ($\chi^2/\mathrm {DOF}=2.8/3$) with a photon index $\mathrm{\Gamma_{\text{intr}} = 1.99 \pm 0.16}$ and a normalization constant $\mathrm{f_{0,\text{intr}}} = (2.03 \pm 0.13) \times 10^{-11}$ \text{ph}\;$ \text{cm}^{-2}\;\text{s}^{-1}\;\text{TeV}^{-1}$  at the same
energy of 200\;GeV. {Other EBL models applied to correct the data provide compatible results.}

Since the resulting SED from 2WHSP~J073326.7+515354 at VHE is substantially flat, the source has a second hump likely peaking at few TeV (see later for further details). This is a first difference with respect to the hard-TeV blazars of \citet{Nustar_EHBLs}, which show continuously increasing flux {up to at least} several TeV and hard spectral index of the order of $1.5 \sim 1.7$. A summary of the source characteristics and results from the VHE data analysis can be found in \Cref{tab:srctable}. {The flux results above 200\,GeV as a function of the observation time are given in \Cref{table-magic-observations}.}


\subsection{\emph{Fermi}-LAT data analysis}
\label{sec:fermi}

The pair-conversion Large Area Telescope (LAT) on board the \emph{Fermi} satellite monitors the gamma-ray sky in survey mode every three hours in the energy range from 20 MeV to $>300$~GeV \citep{Fermitelescope}. 
For this work, a region of interest (ROI) centered around 2WHSP~J073326.7+515354 (4FGL J0733.4+5152) with a radius of 7$^\circ$ was selected. The data sample included more than ten years of data collected by \emph{Fermi}-LAT, from 2008 August 4 to 2019 June 24 (MJD 54682-58658). The data reduction of the events of the Pass8 \texttt{source} class was performed with the Science-Tools software package version v11r5p3 in the energy range 0.5-300\;GeV. To reduce Earth limb contamination a zenith angle cut of 90$^\circ$ was applied to the data. The binned likelihood fit of the data was performed using the recommended Galactic diffuse emission model \citep[see e.g.,][]{2016ApJS..223...26A} and isotropic component recommended for Pass8 (P8R2) \texttt{source} event class\footnotemark\footnotetext{\url{https://fermi.gsfc.nasa.gov/ssc/data/access/lat/BackgroundModels.html}}. 

The normalizations of both diffuse components in the source model were allowed to freely vary during the spectral fitting. In addition to the source of interest, all the sources included in the 4FGL catalogue \citep{2019arXiv190210045T} within a distance of 14 degrees from the source of interest were included. {We build the likelihood model including all the 4FGL sources within 14 degrees from the  position. For the likelihood minimization we leave free to vary the spectral parameters of the sources in the region within 5 degrees from the centre of the RoI and fixed them to their catalogue. values outside. The binned likelihood fit was carried out in two steps, After a first fit, the targets with Test Statistics (TS) $<2$ were removed from the model. After that cut, a final likelihood fit was carried out. We did not find significant residuals, which could suggest the presence of additional sources in the ROI. 2WHSP~J073326.7+515354 was detected with a TS=138.8, a flux of $F(0.5-300 \mathrm{GeV})=(1.3\pm0.5)\times 10^{-9}\; \mathrm{ph}\;\mathrm{cm}^{-2}\,\mathrm{s}^{-1}$ and a hard spectral index of $\Gamma=1.73 \pm 0.11$ (compatible with the value reported in the 4FGL catalogue, $\Gamma_{4FGL}=1.80 \pm 0.10)$. The same analysis is carried out in 2-year time bins to study the flux evolution of the source. The results are shown in \Cref{table-fermi-lc}. }

\begin{figure}
    \centering
    \includegraphics[width=\textwidth]{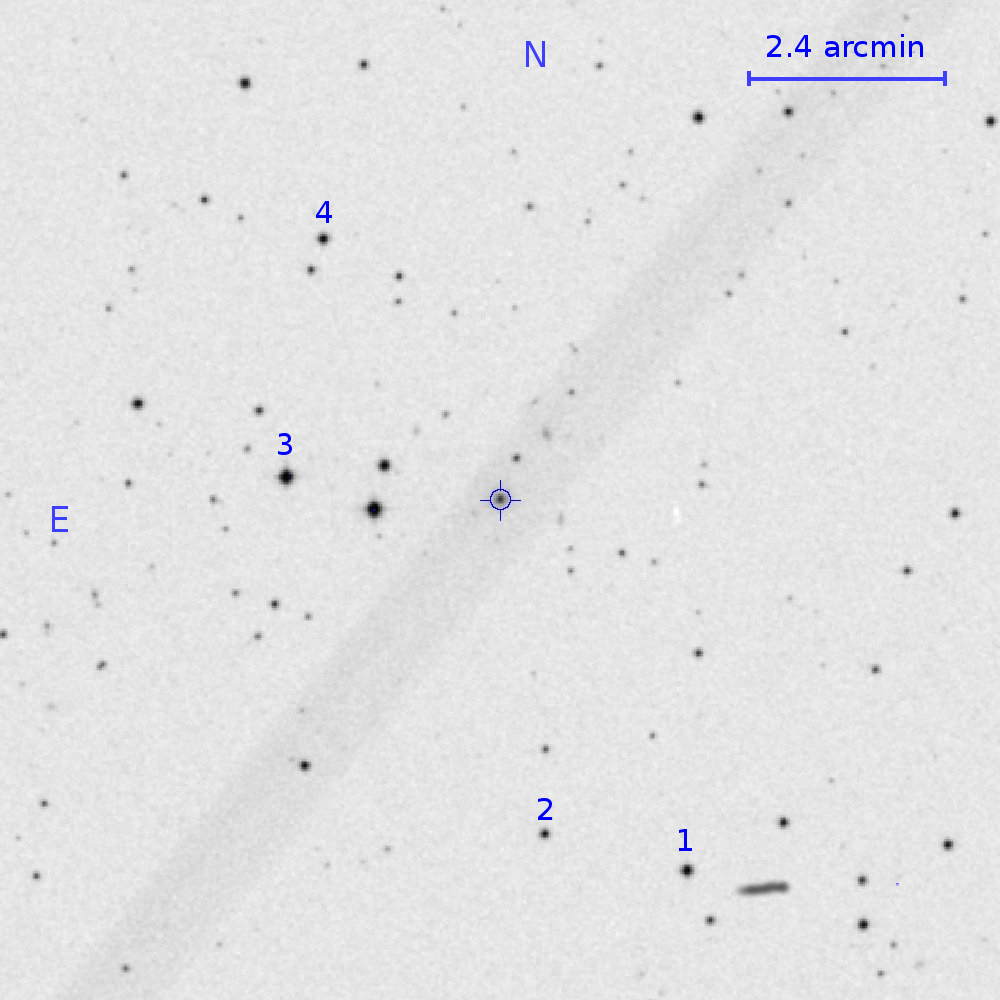}
    \caption{Finding chart of 2WHSP~J073326.7+515354 for the optical photometry and host galaxy measurement. It has been produced from the Digitized Sky Survey (DSS) images using SkyView (The Internet's Virtual Telescope, \url{https://skyview.gsfc.nasa.gov/current/cgi/titlepage.pl} )
    }
    \label{kvafchart}
\end{figure}

\subsection{\emph{Swift} data analysis}
\label{sec:swift}

During the MAGIC observation campaign, simultaneous optical-UV and X-ray observations were performed with the \textit{Neil Gehrels Swift Observatory (Swift)} via a Target of Opportunity (ToO) request.


\begin{figure*}
\centering
{\includegraphics[width=\columnwidth]{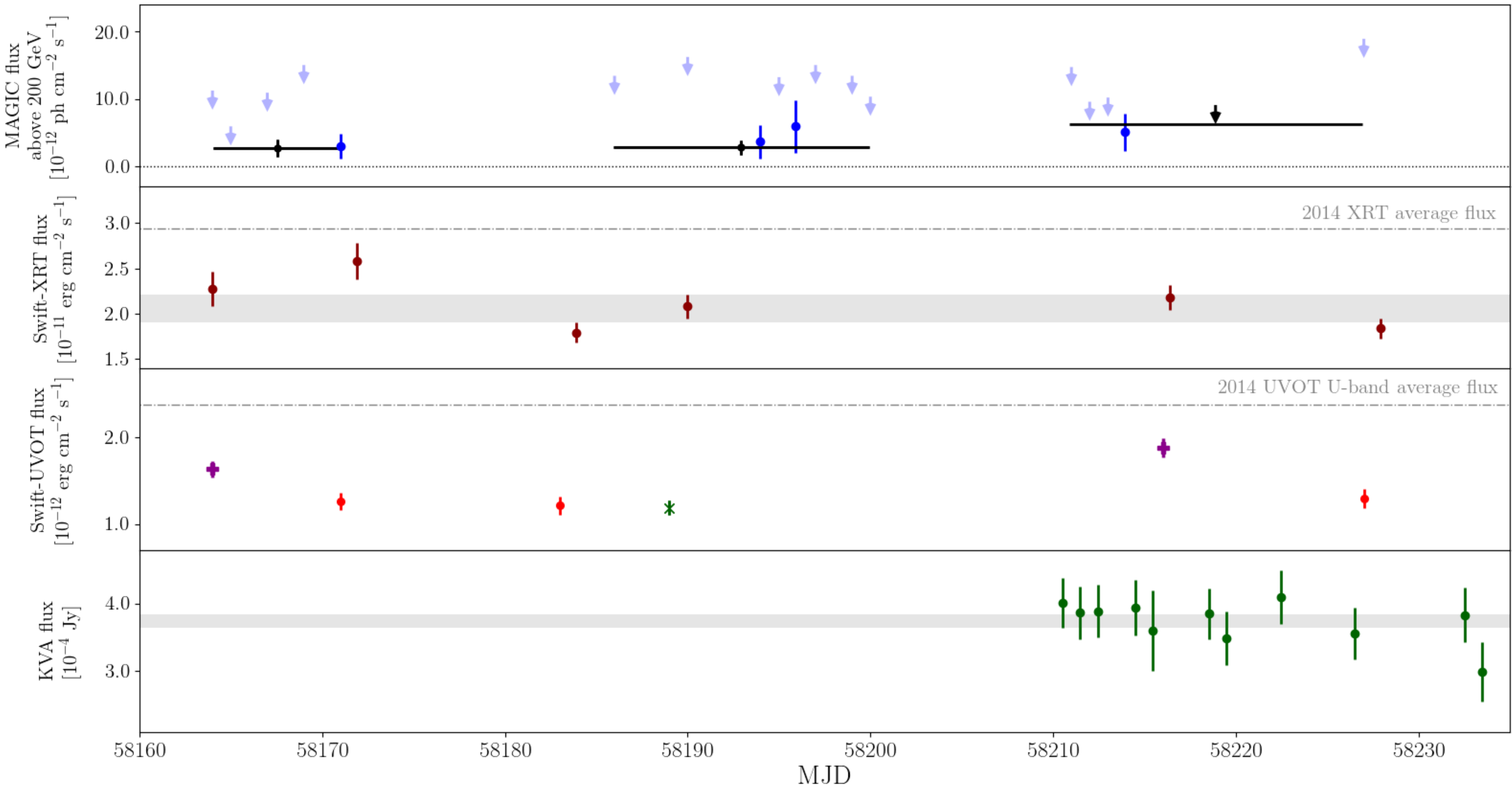}}
\caption{MWL light-curve of 2WHSP~J073326.7+515354 during the MAGIC observation campaign. In order from top to bottom, we present the MAGIC flux and 95\% C.L. upper-limits (arrows) above 200 GeV, the \emph{Swift}-XRT 0.3-10 keV flux points, the \emph{Swift}-UVOT points (in the U-band in violet, W1-band in red, and W2-band in dark green), and the KVA optical flux points (after host galaxy subtraction). For the X-ray and optical energy bands, we report in light grey the 1$\sigma$-band around the average flux. The MAGIC flux has been computed in night-wise (in blue) and monthly binning (black). {For the X-ray and UV energy bands, we report in dashed light grey lines the highest average flux obtained during 2014 observations. Due to the low flux emitted by the source in the HE band, the light-curve as observed by the {\it Fermi}-LAT can only be produced in large time bins larger than the scale shown in this figure.}} 
\label{fig:mwlLC}
\end{figure*}

\subsubsection{XRT instrument}

The X-ray Telescope \citep[\textit{XRT},][]{2004SPIE.5165..201B} on board \textit{Swift} acquired eight good quality raw {datasets}\footnotemark\footnotetext{\url{https://swift.gsfc.nasa.gov/analysis/threads/gen_thread_attfilter.html}}. These eight observations {cover the period between} 2018 January 26 (MJD 58144.08) and 2018 April 29 (MJD 58227.92), and have a total exposure time of $\sim 2.7\;$h with an average of 1.2\,ks per observation. The observation data were analysed based on the standard \textit{Swift} analysis procedure described by \citet{2009MNRAS.397.1177E} using the configuration described by \citet{Ramazani:2017} for the photon counting observation mode and assuming a fixed equivalent Galactic hydrogen column density of $N_H = 5.12 \times 10^{20}$ cm$^{-2}$ \citep{2005A&A...440..775K}.

The spectra for each individual daily observation were fitted by a power-law and a log-parabola function. In all cases, {the \mbox{log-parabola} fit did not improve significantly {the result} (lower than 3 sigma confidence level -C.L.-) with respect to the power-law fit}. The results of this analysis are reported in \Cref{table-swift-observations} together with data obtained by \textit{Swift}-XRT since 2009. The X-ray spectrum of the source is hard, with a photon index  $1.5 \leq \Gamma_X \leq 1.6$  {on the data strictly simultaneous to the MAGIC campaign, but with hint of a softer spectrum in the archival data with larger uncertainty. }

\subsubsection{UVOT observations}

During the {\em Swift} pointings in 2018, the UVOT instrument observed 2WHSP~J073326.7+515354 in its optical (U) and UV (W1 and W2) photometric bands \citep{poole08,breeveld10}. We analysed the data using the
\texttt{uvotsource} task included in the \texttt{HEAsoft} package
(v6.23). Source  counts were extracted from a circular region of 5 arcsec radius centred on the source, while background counts were derived from a circular region of 20 arcsec radius in a nearby source-free region. 
The observed magnitudes are corrected for extinction using the E(B--V) value of 0.50 from \citet{schlafly11} and the extinction laws from \citet{cardelli89} and converted to flux densities. { The results for each individual observation are shown in \Cref{table-swiftUV-observations}.}


\subsection{KVA data analysis}
\label{sec:kva}
The Tuorla blazar monitoring program\footnotemark\footnotetext{\url{http://users.utu.fi/kani/1m}} has observed 2WHSP~J073326.7+515354 coordinated with the MAGIC observations since 2018 April. These observations were performed in the R-band (Cousins) by the 35\,cm telescope attached to the Kungliga Vetenskapsakademien Academy (KVA) system. The data were analysed using the differential photometry method described by \citet{2018A&A...620A.185N}. In order to perform differential photometry, the comparison stars were selected in the same field of view (reported in \Cref{kvafchart}). To measure their magnitude,
the source was observed among many other blazars with known comparison stars on the night of 2018 April 2. The results of the calibration in the R band of the comparison stars in \Cref{kvafchart} are: star n.1  with magnitude 13.11, star n.2  with magnitude 14.29, star n.3  with magnitude 11.93, and star n.4  with magnitude 13.63.
The average zero point of the night was calculated from the photometric zero point magnitude of each image using constant aperture taking into account the effect of airmass.

\begin{figure}
\centering
\includegraphics[width=\columnwidth]{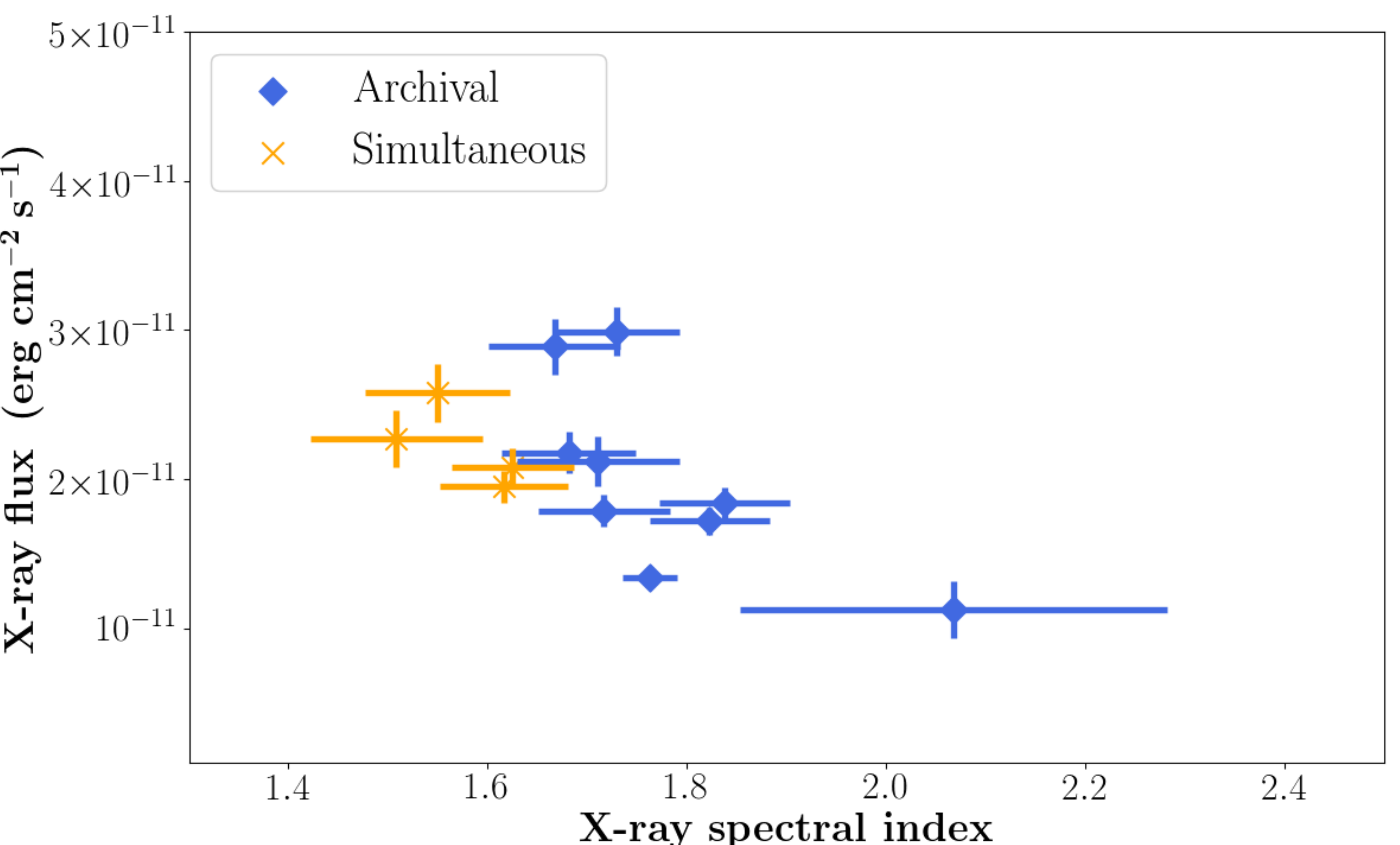}
\caption{The X-ray 0.3-10 keV integral flux as a function of the spectral index measured with {\it Swift}-XRT during all the {previous non-simultaneous} observations of the source (in blue) and the ones simultaneous to MAGIC observations (yellow).}
\label{fig:harderwhenbrighter}
\end{figure}

The contribution of the host galaxy flux is calculated by combining 55 good quality images taken by the KVA telescope.  The images are treated for bias, dark, flat-fielding and fringe map corrections. {A Markov Chain Monte Carlo ensemble sampler \citep[e.g][]{2017arXiv170404629M} was used to map an \textit{a posteriori} distribution in the three-dimensional parameter space.} The resulting images are aligned using the stars in the FoV and the median combining method. The combined image has a total exposure of 5500~s with a full width at half-maximum of FWHM$ =3.0$\arcsec. The comparison star No. 3 (\Cref{kvafchart}) was used to calibrate the field. Following the method described in \citet{2018MNRAS.480..879M}, we use the combined deep R-band image to search for the host galaxy emission.  

In order to study its host galaxy, we fitted two-dimensional surface brightness models to the light distribution of 2WHSP~J073326.7+515354. First, we fit a point source model (with three free parameters, i.e. the source x--y positions and the nucleus flux) with the Sersi\'c index equal to 4 to fix the position of the nucleus. A second fit was performed with a model of a point source and an elliptical host galaxy of ellipticity equal to zero.
The positions determined in the first model were used as first order approximation for the position of the core and the host galaxy. Both models were convolved with the PSF which was determined from the comparison with star n.3. The fits were made to pixels within 18 pix from the center of 2WHSP~J073326.7+515354.  We employed 50 independent walkers, each completing $2 000$ iteration steps and with flat priors. The best-fitting (mode of the posteriors) parameters of the second model are host galaxy flux $R_\text{host}$ = 14.88 mag and effective radius $r_\text{eff}=  6.8\arcsec$. The core flux in the R-band optical is $R_\text{core}$~=~17.36~mag. From these data, the host galaxy flux within an aperture of $5\arcsec$  is $F_\text{host}=1.38$~mJy.


The results of the Tuorla blazar monitoring are presented in  \Cref{tab:kva}. They are corrected for Galactic extinction using values from \citet{schlafly11} and the host galaxy contribution.


\section{Variability}
\label{sec:variability}

The MWL light-curve from optical to VHE gamma rays is shown in \Cref{fig:mwlLC}. The X-ray observations
{allow us to study the variability of the synchrotron flux and the peak of its emission}, while the gamma-ray light-curve from  MAGIC can be used to infer the flux evolution in the high energy peak.
Since the light-curve from \emph{Fermi}-LAT has been computed on large time bins of two years over all ten years of observations, we report  in \Cref{table-fermi-lc} the flux and { photon index} measurements in this energy band.

In the optical band, the KVA observations during the MAGIC campaign were carried out in the R filter. The results are compatible with a constant flux of $3.74 \pm 0.1 \; \text{mJy}$, yielding a {$\chi^2$/DOF of 6.6/11. }

In the UV band, as observed by {\it Swift}-UVOT with filters U, W1 and W2, even though the statistics are sparse, no strong flux variations were detected over the course of the MAGIC observation campaign. The flux is compatible with a constant fit of $(1.3 \pm 0.2) \times 10^{-12}\;\text{erg cm}^{-2}\,\text{s}^{-1}$ with {$\chi^2$/DOF of 0.5/3} and $(1.76 \pm 0.05)  \times 10^{-12}\;\text{erg cm}^{-2}\,\text{s}^{-1}$ with {$\chi^2$/DOF of 4.6/4} for the bands W1 and U, respectively. For the band W2, only one observation is available during that period, and therefore no conclusion for variability can be derived. As reported in \Cref{table-swiftUV-observations}, in comparison with historical observations from 2009  and 2011, the source shows fluxes compatible with the average fluxes reported above. However, during January 2014  the source showed {fluxes higher  by a factor of  about 4-5 times compared to} the average flux during the MAGIC observation window in both the U and W2 bands.

The X-ray observations performed by {\it Swift}-XRT during the MAGIC observation campaign show moderate variability. A constant fit to the flux evolution during that period can be discarded at a 3.7 $\sigma$ C.L. ({$\chi^2$/DOF of 29.6/8}). The previous observations of the source carried out between 2009 and 2014 show a flux range of 1 to $3 \times 10^{-12}$ $\text{erg cm}^{-2}\;\text{s}^{-1}$. When considering only simultaneous XRT and MAGIC observations (MJD 58144, 58164, 58190, and 58227), the flux is compatible with a constant average flux of ($2.07 \pm 0.15) \times 10^{-11}$ \text{erg}\;$ \text{cm}^{-2}\;\text{s}^{-1}$ {($\chi^2$/DOF of 6.8/4)}. A marginal ``harder-when-brighter'' trend is found in the flux vs spectral index observed in the X-ray band by \textit{Swift}, as shown in \Cref{fig:harderwhenbrighter}. The trend can be fitted by a linear function with {$\chi^2$/DOF of 1.3/13)} with slope of $-1.64 \pm 0.62$. This trend is quite typical in {BL~Lacs}, and has been observed in several X-ray campaigns of Mrk~501 and Mrk~421 \cite[e.g.,][]{harder-when-brighter,mkr501-flare2012}.

Finally, in the {\it Swift}-BAT 105-months catalogue \citep{BATcatalog105} the source is detected with a signal to noise ratio (SNR) of only 5.38, and no variability is reported.

Regarding the high-energy SED peak, the light curve is limited due to the low flux of the target. For the flux evolution of the HE gamma rays observed by \textit{Fermi}-LAT, a time bin of two years was used in order to collect enough photon statistics. As shown in \Cref{table-fermi-lc}, due to the weak detection,
the measured flux is compatible with a constant flux during the first 10 years of operation of \mbox{\textit{Fermi}-LAT}. 
{We checked the possible enhanced flux of the source of interest in the period around January 2014, when there was an optical-UV flux registered by \emph{Swift}-UVOT few times higher with respect to the average flux measured during the MAGIC observation window. An analysis of the \emph{Fermi}-LAT data over the period from August 3$^\text{rd}$, 2012 to August 3$^\text{rd}$, 2014 (MJD 56142.7- 56872.7, as reported in \Cref{table-fermi-lc}), which includes January 2014, reports no photons detected with probability >50\% of belonging to the source of interest.}
This result is also compatible with the variability index of 39 reported in the \emph{Fermi}-LAT 3FGL catalogue \citep{fermi3fgl}, {statistically consistent} with a steady source (variability threshold 72.44 { as reported in the 3FGL catalogue)}. Moreover, the long-term SED measured by \mbox{\textit{Fermi}-LAT} connects smoothly with the VHE gamma-ray SED observed by MAGIC as shown in \Cref{fig:mwlSED}. 
Thus, while short term variability cannot be excluded due to the low photon statistics and the fact that the long integration of the signal might smooth out some modest flux variations, the stability of the light curve on the long term supports a steady flux condition of the source within the sensitivity of the instrument.

For the VHE band, as shown in \Cref{fig:mwlLC} the source is detected above $2\sigma$ C.L. only during four nightly observations (blue points and arrows). The rest of the observations yield upper limits. Due to the lack of strong variability detected from the nightly observations {(constant fit with $\chi^2$/DOF of 16/21)} and the low photon statistics, the monthly light-curve is also evaluated (black points). 
{Also with larger time bins, the light-curve does not show any hint of variability, and the average flux results in $(3.4 \pm 0.4) \times 10^{-12} \;\text{ph cm}^{-2}\,\text{s}^{-1}$ with {$\chi^2$/DOF of 1.2/4. }
}

In summary, when considering only the MWL data simultaneous to the MAGIC observations, no significant variability is identified. Therefore, during the MAGIC observation campaign the source remained in a stable state. Only some moderate variability was { measured by {\it Swift}-UVOT/XRT} when comparing with historical observations.

\section{Multi-wavelength SED} 
\label{sec:sed}

We present in \Cref{fig:mwlSED} the SED {with the full data sample} we assembled for 2WHSP~J073326.7+515354. In grey, we show the selected archival SSDC data (see caption for details). Then we report in orange the KVA data, in blue the  \mbox{\emph{Swift}-UVOT} data, in light red   the  \emph{Swift}-XRT data, in dark red   the  \emph{Swift}-BAT data, in purple the \emph{Fermi}-LAT data, and in dark green the MAGIC data. In order to account for the modest variability found with the \emph{Swift} data, we will consider only its data strictly simultaneous to MAGIC observations (MJD 58144, 58164, 58190, and 58227).

Since the synchrotron peak position $\nu_{\text{peak}}^{\text{sync}}$ is the basis of the definition of EHBL, this value plays an important role in classifying new sources of this class.
In order to measure $\nu_{\text{peak}}^{\text{sync}}$, we performed a log-parabolic fit of the synchrotron peak of 2WHSP~J073326.7+515354 {as illustrated in \Cref{fig:synchrotronpeak}}. The fit performed only on the \emph{Swift}-XRT  \mbox{X-ray} data simultaneous to MAGIC observations, being compatible with a power-law model, does not allow us to constrain the synchrotron peak location. For this reason, the non-simultaneous \mbox{\emph{Swift}-BAT} {105-month} archival data were used to provide a first estimation of the synchrotron peak. 
The resulting new estimation leads to $\nu_{\text{peak}}^{\text{sync}} \simeq {10^{17.8 \pm 0.3}}$ Hz {($\chi^2$/DOF of 19/39)}. This value is compatible with the estimation reported in the 2WHSP catalogue \citep{2whsp} of \mbox{$\nu_{\text{peak,2WHSP}}^{\text{sync}}=10^{17.9}$~Hz} and confirms the classification of 2WHSP~J073326.7+515354 as an EHBL.

Due to the high-frequency location of the synchrotron peak, the SED exhibits the optical radiation of the host galaxy. The combination of the simultaneous KVA and \emph{Swift}-UVOT data allows us to build a good template for the host galaxy radiation in the optical range of the SED. 

Thanks to the set of MAGIC and \emph{Fermi}-LAT SED points, we are now able to
study the IC peak $\nu_{\text{peak}}^{\text{IC}}$, reported in Figure~\ref{fig:ICpeak}.
In the HE gamma-ray band, the \emph{Fermi}-LAT points present a hard spectral index of $\Gamma_{HE}=1.67 \pm 0.11$. This means that {they constitute}
the rising part of the second SED hump that finally peaks in the TeV gamma-ray band. {For this reason, given the hard gamma-ray spectrum, we fitted the (EBL-deabsorbed) second hump with a \mbox{power-law} model, reporting a {($\chi^2$/DOF of 5.2/7)} and a slope of $2.13 \pm 0.04$.
Alternatively, the EBL-deabsorbed spectrum can be fitted also with a \mbox{power-law} model with exponential cut-off {($\chi^2$/DOF of 4/8)}. 
This allows for an estimation of the cut-off}  \mbox{$\nu_{\text{cutoff}}^{\text{IC}}  10^{27.2\pm 0.2}$ Hz},
{and thus that the second SED hump peaks at $\nu_{\text{peak}}^{\text{IC}} = 10^{26.4 \pm 0.6}$ Hz.}

The Compton dominance (CD) parameter for the different models, reported in \Cref{tab:modeling}, is defined as {the ratio between the second hump peak luminosity and the synchrotron peak luminosity $\nu L_\nu$. }  
Considering our estimation of the two SED peaks, this parameter results in \mbox{$CD\sim0.12$}. This result is compatible with the {phenomenological} CD trend observed for the gamma-ray blazar sample detected by \emph{Fermi}-LAT reported in \cite{finke}: the higher the frequency of the synchrotron peak the lower the CD value. The low value for the CD parameter agrees with the conventional interpretation of poor environments {without strong low energy photon fields} around the EHBL relativistic jets preventing the high-energy emission via EC scattering (contrary to the rich external fields in FSRQs for instance).

\section{Modeling}
\label{sec:modeling}

Four different emission models have been tested on the experimental data for the emission of the blazar jet. First, we start with the application of two { different} one-zone leptonic models. Testing such models, we face the need for applying {extremely low magnetization} within the emission region. Therefore, we use two different approaches to try to overcome this problem: a two-zone leptonic model and a hadronic model. 

{In addition, the template for a typical host galaxy contribution is applied to the broad-band SED following \cite{hostgalaxymodel},} adapted to the redshift of 2WHSP~J073326.7+515354 (z=0.065). {It is worth to note that this model represents only a reference model for the host galaxy emission, and is not fitted to the data. Differences between the model and some  archival data might be due to different apertures adopted by the different instruments, or specific observing conditions, but no detailed information about this is available in the database.


\newpage
\thispagestyle{empty}
\begin{landscape}
\begin{figure}
\centering
\subfloat[SSC model.]{\includegraphics[width=0.49\textwidth]{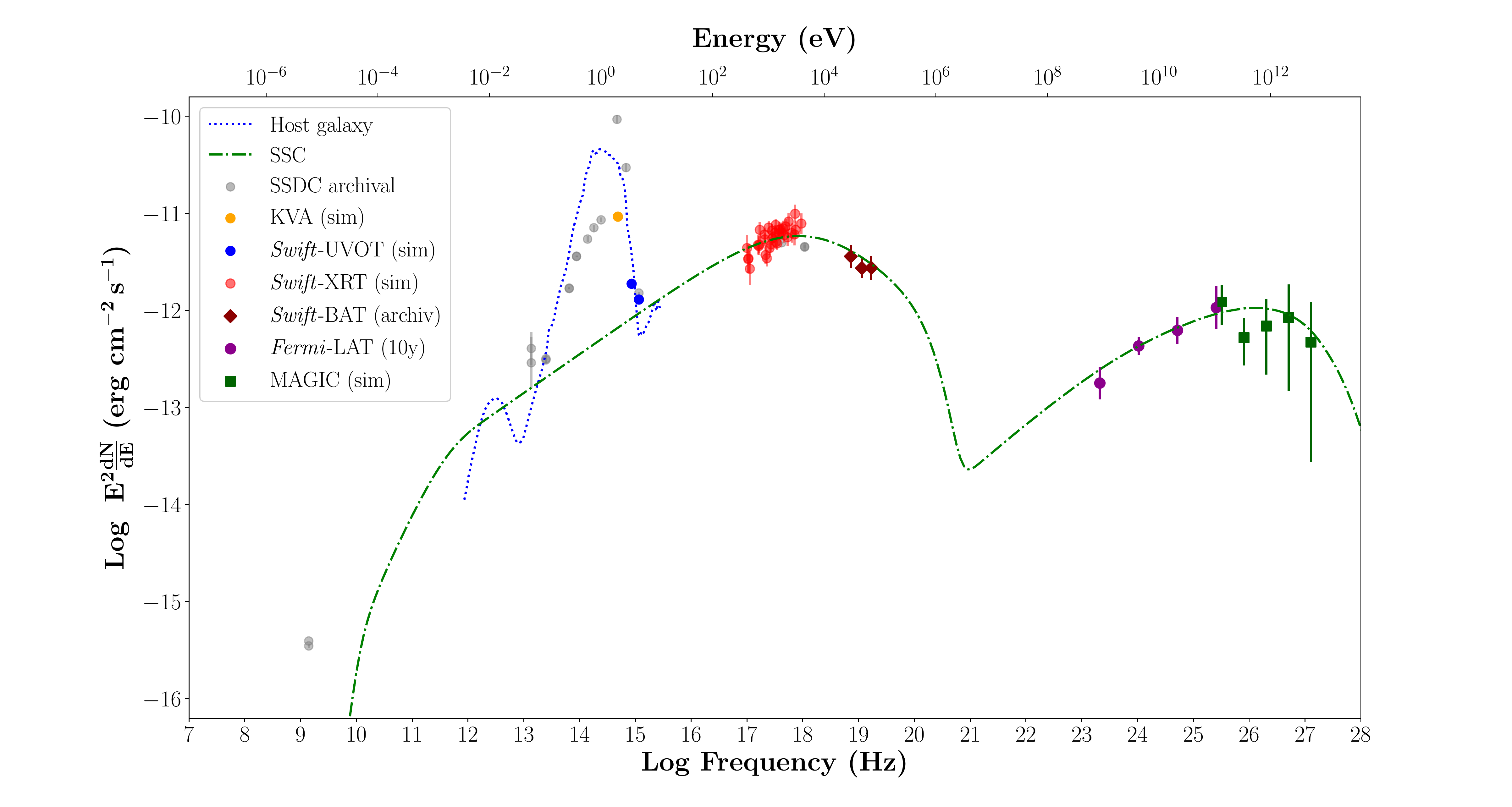}\label{fig:modelingSED-SSC}
}
\subfloat[1D conical jet model.]{\includegraphics[width=0.49
\textwidth]{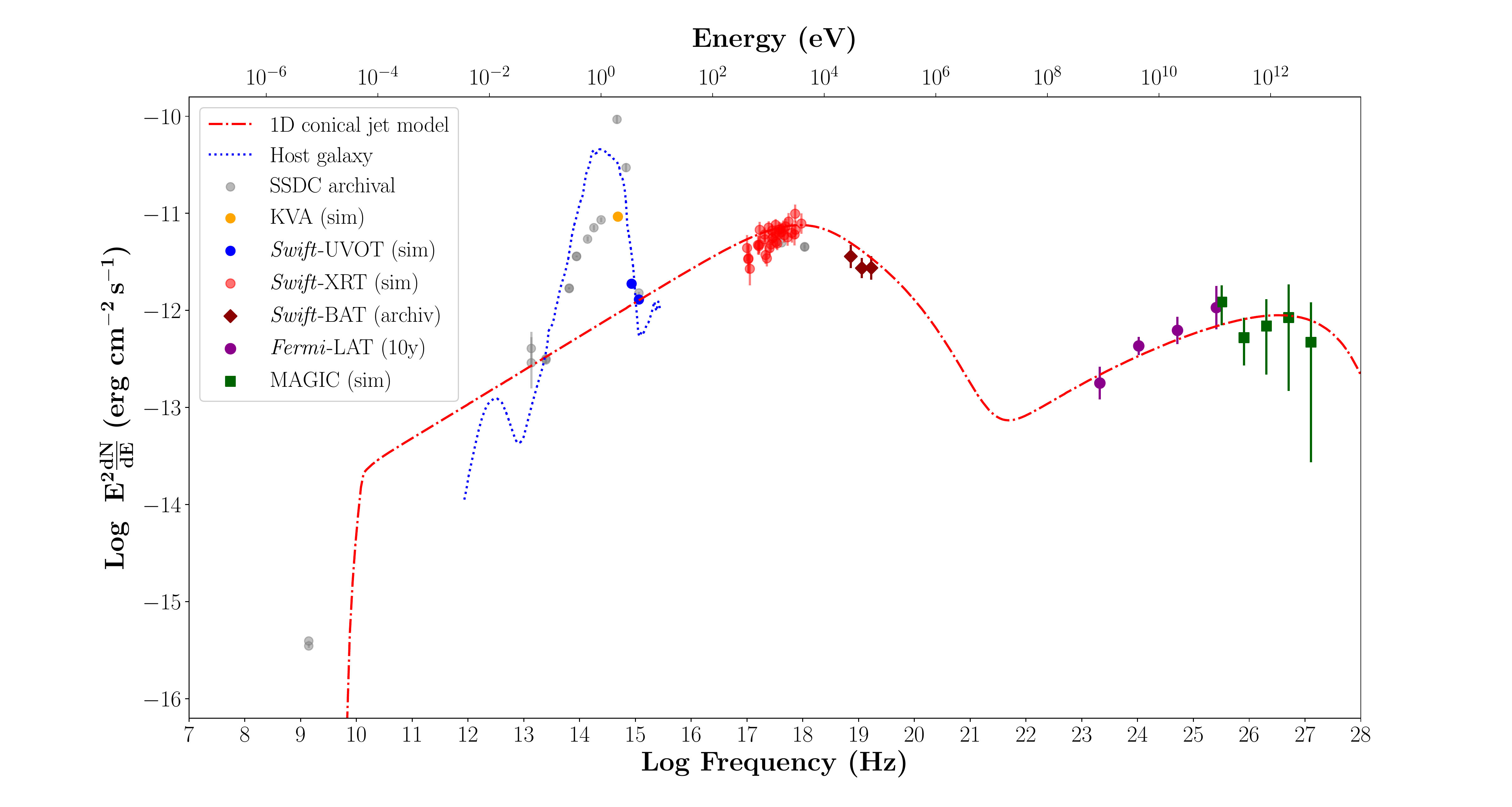}\label{fig:modelingSED}
}\\
\subfloat[Spine-layer model.]{\includegraphics[width=0.45\textwidth]{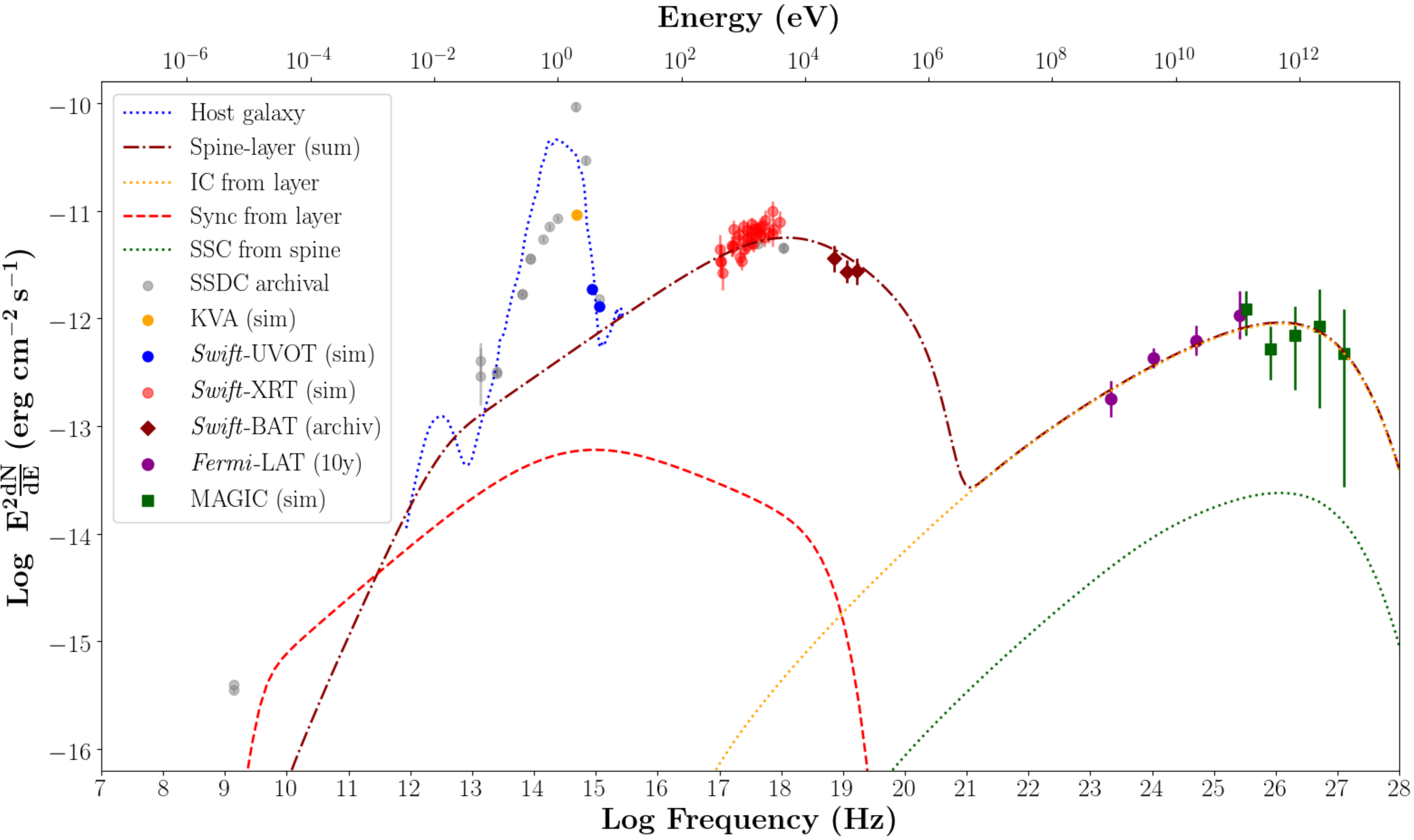}\label{fig:modelingSED-spinelayer}}
\subfloat[Hadronic model. {Here we report all the models that survived the $\chi^2$ cuts.}]{\includegraphics[width=0.45\textwidth]{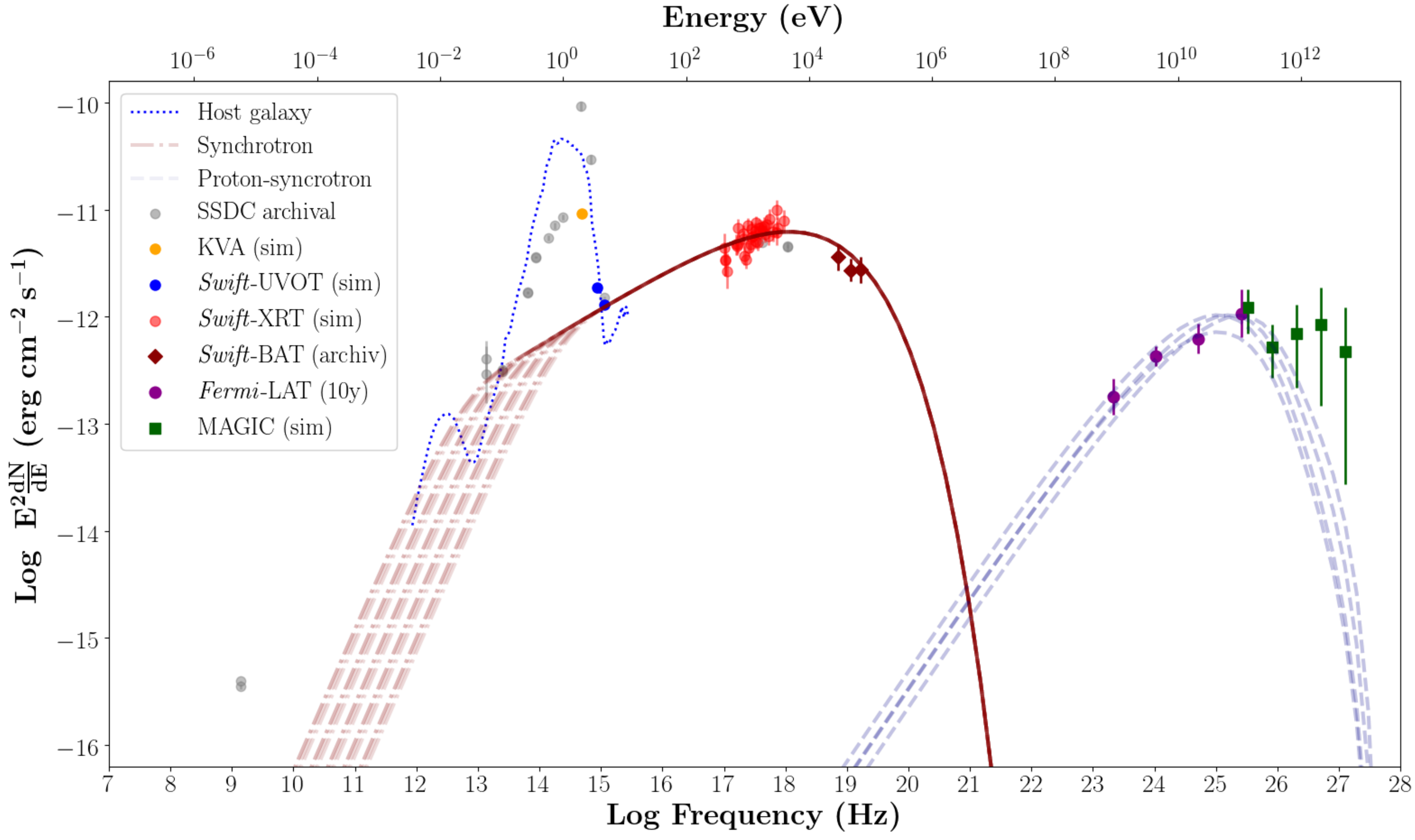}\label{fig:hadronic-model}}
\caption{MWL intrinsic SED of 2WHSP~J073326.7+515354. The data have been EBL de-absorbed using the model by \citet{Dominguez:2010bv}.  In grey, we report the selected archival SSDC data (\emph{First} data on MJD 49078 in \citet{first-data-ssdc}, \emph{WISE} data between MJD 55287-55479), in orange the KVA data, in light red the  \emph{Swift}-XRT data, in dark red the  \emph{Swift}-BAT data, in purple the \emph{Fermi}-LAT data, and in dark green the MAGIC data. Arrows represent upper-limits.}
\label{fig:mwlSED}
\end{figure}
\end{landscape}
\restoregeometry
\newpage


\noindent
Any conclusion based on the comparison between the model and the data would require much more careful analysis beyond our scope.}

{Given that no significant variability is observed in the source (see Section 3), we fix the emitting region size to $R=10^{16}$~cm (a typical value for HBLs).}

Finally, considering that electrons and positrons cannot be distinguished from a radiative perspective, we will refer to both populations as electrons.

\subsection{One-zone leptonic models}

\subsubsection{Synchrotron self-Compton model} 
\label{SSC}

The synchrotron self-Compton model is the standard \mbox{one-zone} leptonic model historically used to model the MWL emission of BL~Lac type objects \cite[e.g.,][]{Maraschi:SSC,Tavecchio:SSC}. In this scenario, the emission is produced by relativistic electrons contained in a spherical region of {radius $R = 10^{16}$ cm} with a tangled and uniform magnetic field $B$. This region is moving with a bulk Lorentz factor $\Gamma$ along the axis of a relativistic jet, which forms an angle $\theta$ with respect to the observer line of sight.
The special relativistic effects are accounted for by the relativistic Doppler factor $\delta=[\Gamma(1-\beta \cos \theta)]^{-1}$. The model assumes the presence of a population of relativistic electrons of density $N$ distributed with a broken power-law spectrum as a function of the {Lorentz factor} of the electrons:
\[
\hspace{60pt}N(\gamma)=\,K \,\gamma^{-p_1}\, \Big(1+\frac{\gamma}{\gamma_b}\Big)^{p_1-p_2}\; ,
\]
where $K$ is the normalization factor, and $p_1$ and $p_2$ are the spectral indices respectively before and after the spectral break, at which the Lorentz factor of the electrons is $\gamma_b$.

{Electrons produce synchrotron radiation that is in turn Compton-scattered generating the high-energy SSC continuum.}
As detailed in \citet{Tavecchio:SSC}, this simple model is fully constrained if a good sampling of the SED (especially around the peaks) and an estimate of the variability timescale are available. The good dataset collected for 2WHSP~J073326.7+515354 is therefore quite suitable {for the application of this model in \Cref{fig:modelingSED-SSC} and provides} strong constraints on the physical parameters of the jet, {reported in \Cref{tab:modeling}.}

{By applying the SSC model to the data, we can provide an estimation of the synchrotron peak located at $\nu_{\text{peak}}^{\text{sync}} =10^{18}$~Hz ($\simeq 4.0$~keV) and an IC peak located at $\nu_{\text{peak}}^{\text{IC}} =10^{26.4}$~Hz \mbox{($\simeq 1.2$ TeV)}. These values are in agreement with the observational fits we reported in Section~\ref{sec:sed}.
}

\subsubsection{1D conical jet model} 

The MWL SED has been modeled also adopting the numerical code by \citet{2014ApJ...780...64A} \citep[see also][]{2015ApJ...808L..18A,2018arXiv180509953A}, reporting the results in \Cref{fig:modelingSED}. This model calculates the emission from the non-thermal electrons in a conical jet. The evolution of the electron and the photon energy distributions are followed along the motion of the jet. This framework is similar to the {\tt BLAZAR} code by \citet{2003A&A...406..855M},
which has been frequently adopted to reproduce blazar spectra \citep[see e.g.,,][]{2008ApJ...672..787K,2012ApJ...754..114H}.

The conical expansion of the jet naturally leads to adiabatic cooling of the electrons. This effect resembles the electron escape in one-zone steady models, that can thus be neglected in this 1-D~code.
The model assumes a continuous injection of non-thermal electrons from
the initial radius $R=R_0$ during the dynamical timescale $R_0/c\Gamma$ in the plasma rest frame.  In this timescale, the injection rate into a given volume $V$ - which is expanding as $V \propto R^2$ - is assumed to be constant.
The magnetic field $B$ in the plasma frame evolves as $B$ = $B_0 \, (R_0/R)$.
We take into account the synchrotron and inverse Compton scattering with the Klein-Nishina effect, the  $\gamma$-$\gamma$-absorption, the secondary pair injection, the synchrotron self-absorption, and the adiabatic cooling.

The electron energy distribution at injection is assumed as a broken power-law
with exponential cut-off, where the parameters are low-energy index $p_1$,
high-energy index $p_2$, and the break energy (Lorentz factor) $\gamma_{\rm br}$
and the cut-off energy $\gamma_{\rm max}$.
The minimum Lorentz factor $\gamma_{\rm min}$ is fixed as 20.
The electron energy distribution and the photon emission are computed even after electron injection {ends, until} $R$ reaches $R$ = 10 $R_0$.

\begin{table*}
\renewcommand{\arraystretch}{1.37} 
\scalebox{0.8}{
\centering
	\hspace*{-25pt}
		\begin{tabular}{lcccccccccccccc}
			\hline\hline
		Model	  & Component & 
		$\Gamma$ & K &  $B$ &  $R$ ($R_0$/$\Gamma$ in 1D SSC model) & $L_{\rm e}$  & $\gamma_{\rm min}$& $\gamma_{\rm br}$ & $\gamma_{\rm max}$ & $p_1$ & $p_2$  & CD  & $U_B/U_{\rm e}$ \\
			 &  && &  G &$\mbox{cm}$ & $\mbox{erg}~\mbox{s}^{-1}$ & & &  &  & && \\
			\hline
			One-zone SSC & & $30$ &$7.7  \times 10^{3} $& 0.01 &$1 \times 10^{16}$&  $6.2 \times 10^{43}$ & 500 &$ 1 \times 10^{6}$ & $1 \times 10^{7}$ & 2.2 & 4.0    & 0.12 & $5 \times 10^{-4}$\\
		    1D SSC&& $30$ &  & 0.005 & $2.1 \times 10^{16}$ &  $1.2 \times 10^{45}$ & $20$   &$ 2 \times 10^{6}$ & $2 \times 10^{7}$ & 2.3 & 3.5  & 0.12 & $7 \times 10^{-5}$\\
		  Spine-layer & spine & $30$ & $7.5 \times 10^{1} $& 0.02 &  $3 \times 10^{16}$ & $1.2 \times 10^{45}$ & 1000 & $ 9 \times 10^{5}$ & $8 \times 10^{6}$ & 2.2 & 4.1    & 0.14 & 0.26\\
		 & layer & 5 &  $1 \times 10^{1}$ & 0.1 &  $3.5 \times 10^{16}$ &  & 1 & $1 \times 10^{4}$ & $3 \times 10^{6}$ &2& 3.5  &  & \\
	     		\hline\hline
		\end{tabular}
}
\caption{Resulting values of the parameters used by the  three leptonic models in the paper. We report the bulk Lorentz factor $\Gamma$, the magnetic field $B$, and the electron luminosity $L_e$. The electron distribution is assumed to have an index of $p_1$ between $\gamma_{\text{min}}$ and $\gamma_{\text{br}}$, and an index of $p_2$ up to the maximum $\gamma_{\text{max}}$.  Then, we report the Compton dominance parameter CD (the ratio of $\nu L_\nu$  at $\varepsilon_{syn,pk}$ to that at
$\varepsilon_{IC, pk}$), and the energy density ratio of the magnetic field to that of the non-thermal electron distribution ($U_B / U_e$) at the radius where the electron injection terminates. }
\label{tab:modeling}
\end{table*}

In this paper, considering an on-axis observer (viewing angle $\theta_v$ is zero), the jet opening angle is assumed to be 1/$\Gamma$, where $\Gamma$ is the bulk Lorentz factor of the jet. The photon flux is obtained by integrating the emission over the entire jet, taking into account the Doppler boosting by the conically outflowing emission region.

For the steady emission scenario, this model includes eight parameters: the initial radius $R_0$, the bulk Lorentz factor $\Gamma$, the initial magnetic field $B_0$, the electron Luminosity $L_{\rm e}$, $p_1$, $p_2$, $\gamma_{\rm br}$, and $\gamma_{\rm max}$. 
The results are summarized in \Cref{tab:modeling}. The model provides a particularly low magnetic field $B$, and this puts the object far from the equipartition limit by more than five orders of magnitude.

The modeling provides also an estimated synchrotron peak frequency $\nu_{\text{peak}}^{\text{sync}} = 10^{18}$ Hz ($ \simeq 4.0$ keV) and an IC peak frequency $\nu_{\text{peak}}^{\text{IC}} = 10^{26.5}$ Hz ($ \simeq 1.25$ TeV). These values are well in agreement with the observational fits we reported in Section~\ref{sec:sed}.


\subsection{The energy equipartition issue}

In \Cref{tab:modeling}, we summarize the resulting parameters for the first two one-zone leptonic models we used. An inspection of the table shows that the parameters are quite similar in both the SSC model and the 1D conical jet model. 
Both the models present a low magnetic field $B$ of the order of $10^{-2}$ G. The low magnetic field - together with a relatively large Doppler factor - is generally required by the SSC modeling in order to account for the large separation between the two SED peaks.
{Since the total electron energy for $p_1>2$
is dominated by low-energy electrons,
a harder electron spectrum ($p_1<2$) well below the energy responsible for
the synchrotron peak would make it close to equipartition.
A stochastic acceleration model (e.g., \citealt{Asano:2013}) can generate
such a hard spectrum.
\citet{dermer2015} succeeded in reproducing a relation between
the spectral index and the peak Compton frequency in blazars
with an equipartition model adopting log-parabola electron energy distribution motivated by the stochastic acceleration model.
Even with their model, however, the broad-band spectrum of Mrk~501
requires a low magnetization.
It may be difficult to make it close to the equipartition
for EHBLs even with stochastic acceleration models
within \mbox{one-one} or 1D SSC picture (see \citealt{2015ApJ...808L..18A}).}

Another common feature of the modeling results is the high value of the minimum energy of the electrons $\gamma_{\text{min}}$ \citep{Katarzynski2006}.
{The extreme values obtained by our models for these parameters  might be in tension with those commonly adopted to describe standard HBL via SSC emission. Their values are instead in agreement with the ones commonly required in the modelling of the extreme counterpart of this class of sources, the EHBL objects}
(\citealt{Tavecchio:2009zb}, but see also \citealt{cerruti2015} for a comparison of some results on the modeling of other known EHBLs).

In the one-zone leptonic framework, both the standard stationary one-zone model and the 1D conical jet  model imply an extremely low magnetic field $B$ of the emitting region that results in a low ratio of the energy densities  $U_B / U_e$, being far from the equipartition limit. Such conditions are particularly interesting considering that they are not related to flaring episodes of the source, but to their relatively quiescent observed emission. This means that, in the case of leptonic scenarios, a mechanism is expected to continuously keep the emission out of equipartition.

In order to increase the $U_B / U_e$ ratio, a solution is to decrease the number of radiating electrons and increase the minimum Lorentz factor $\gamma_{\text{min}}$. Another way would be to modify the size of the emitting region. 
A larger size of the emitting region $R$ and smaller Doppler factors $\delta$ may be an alternative solution, but these \textit{ad-hoc} values 
would not help enough in bringing the conditions much closer to equipartition, accounting only for few times closer, and not orders of magnitude (e.g., \citealt{Nustar_EHBLs}).

{Alternative solutions that require less extreme parameters are} proposed in the following sections: a spine-layer structured jet and a hadronic model.


\subsection{Two-zone model}
\subsubsection{Spine-layer model}

As shown above, the one-zone models applied to the data of 2WHSP~J073326.7+515354 suggest that the emission region is quite far from equipartition, with the electron energy density dominating over that of the magnetic field by more than 3 orders of magnitude. As discussed by \citet{TavecchioGhisellini16}, this result is commonly found for one-zone models of high-energy emitting BL~Lacs (see also \citealt{InoueTanaka16}). The same authors show that a possible solution allowing equipartition conditions for the emitting region is the spine-layer model introduced by \citet{Ghisellini05}. In this framework, the relativistic jet is supposed to be structured, as suggested by several theoretical and observational hints (e.g., \citealt{TavecchioGhisellini15} for details). Besides the emission from 
blazars, the scenario can satisfactorily reproduce the emission of radiogalaxies \citep{TavecchioGhisellini08}, and can potentially account for the neutrino production in BL~Lac objects \citep{Ansoldi18}.

Specifically, the jet is supposed to consist of two components: a central fast spine  and a slower layer around it. The former moves with Lorentz factor $\Gamma_{\text{spine}} = 10 - 20$ in the inner part of the cylindrical jet, while the external layer has $\Gamma_{\text{layer}} = 2 - 5$ \citep{Nappo17}. The layer acts as a source of soft (synchrotron) photons that, thanks to boosting due to the relative motion of the two jet components, can dominate the radiation energy density in the frame of the spine. In these conditions, the inverse Compton emission from the spine is dominated by the scattering of the layer radiation field, while the SSC component is expected to provide a minor contribution. Due to the larger Lorentz factor, at small angles (such as those characterizing blazars) the spine emission prevails over that of the layer.

In \Cref{fig:modelingSED-spinelayer}, we report the model obtained within the spine-layer framework. The IC emission of the spine is largely dominated by the scattering of the layer radiation field. The parameters for the spine (analogous to those describing the one-zone model above) are reported in \Cref{tab:modeling}.

The introduction of a more complex structure of the jet allows a more relaxed fit of the physical parameters.
The spine-layer scheme offers an increased energy density of the radiation field (supposed to be dominated by the radiation provided by the layer), and this allows us to lower the electron density needed to produce a given IC luminosity.
In turn, in order to account for the reduced number of electrons, the model slightly increases the magnetic field $B$ to keep the same synchrotron luminosity. This fact increases the previous $U_B / U_e$ ratio and brings it closer the equipartition limit, reconciling it with the theoretical expectations. 
{Also adopting lower values of $\gamma_{\text{min}}$ confirms our results concerning the energy densities close to equipartition in this model: for example, $\gamma_{\text{min}}=10$ implies $U_B / U_e = 0.1$, that still represents an acceptable value.}

{The structured jet model provides more appropriate physical conditions closer to equipartition,} and leads to an interesting comparison with its application to the sample of hard-TeV blazars reported in \citet{Nustar_EHBLs}.
In fact, the spine-layer model can provide larger magnetic field that generates a more efficient cooling of the TeV electrons, providing a softer spectrum at TeV energies that { does not agree with the hard spectrum up to several TeV of the  hard-TeV blazars seen in \citet{Nustar_EHBLs}.}

However, such a softer spectrum at TeV energies is observed
in more ``standard'' HBL-like EHBLs, like for example 2WHSP~J073326.7+515354.
This source, according to the results in gamma rays previously found, is an EHBL object that presents
an IC peak well detectable at few TeV. This implies that the \mbox{spine-layer} model is still able to fit the SED (especially the IC peak). Thus, the case of 2WHSP~J073326.7+515354 might be a limit-case in the EHBL population with a high synchrotron peak but not extremely hard TeV gamma-ray spectrum.
This result confirms the great difference that might be hidden in the EHBL population due to different spectral properties in the TeV gamma-ray band. Further observations at TeV energies will be able to increase the statistics and characterize new objects of this population.

{Applying this model to the data, it provides the synchrotron peak located at $\nu_{\text{peak}}^{\text{sync}} =10^{18}$ Hz ($\simeq 4.1$ keV) and an IC peak located at $\nu_{\text{peak}}^{\text{IC}}  =10^{26.5}$ Hz ($ \simeq 1.25$ TeV). These values are in good agreement with the observational fits we reported in Section~\ref{sec:sed}.}

\subsection{Hadronic model}

Another solution in order to interpret the SED of 2WHSP~J073326.7+515354 without using extreme { physical} parameters can be found considering a hadronic origin for the gamma-ray component. Blazar hadronic models, in which the gamma-ray component is ascribed to emission by protons in the jet, or by secondary leptons produced in p-gamma interactions, have been widely studied \cite[e.g.,][]{Mannheim:1993,Aharonian:2000pv,Mucke:2001,Boettcher:2013wxa} as an alternative to leptonic models. One of the major drawbacks of this scenario is that it often requires a high proton power, well above the Eddington luminosity of the black hole powering the blazar. For the particular case of EHBLs on the other hand, due to their relatively low luminosity compared to other blazar subclasses, a successful hadronic modeling can be achieved with an acceptable energy budget \cite[see][]{cerruti2015}. In addition, the absence of fast flares from EHBLs (in contrast with the $\gamma$-ray variability seen in more common HBLs) is also consistent with the cooling time-scales of protons in the jet.  With this scenario in mind, we test a simple proton synchrotron model for 2WHSP~J073326.7+515354, using the numerical code described in \citet{cerruti2015}. We make the following assumptions to reduce the number of free parameters to study: 
\begin{itemize}
\item electrons and protons share the same acceleration process, and thus the index  $\alpha$ of the injected energy distribution is identical;
\item {the maximum proton Lorentz factor $\gamma_{p, max}$ is constrained by the equality of cooling and acceleration time-scales; for the parameters used in the model, the fastest cooling time-scale for protons is the adiabatic one;}
\item { the electron energy distribution at equilibrium is calculated assuming that electrons are cooled primarily by synchrotron radiation;}
\item the emitting region size $R$ is limited by the variability 
    time scale, which is considered to be two days
\item the Doppler factor $\delta$ of the emitting region is fixed to 30.
\end{itemize}
Under these assumptions, we produce 350 hadronic models scanning the following parameter space: the radius of the emitting region $R \in [10^{14} \textrm{cm} -1.46\times10^{17}$ cm], the proton peak synchrotron frequency $\nu_{p,syn} \in [4\times10^{24} - 4\times10^{26}]$ Hz, and the proton normalization $K_p \in [K^\star / 3, 3 K^\star]$, where $K^\star$ corresponds to the proton density such that the peak of the proton synchrotron component is at the level of the MAGIC data. A $\chi^2$ test is used to select the solutions which correspond to a 1 $\sigma$ C.L., {obtaining a best $\chi^2$/DOF of 46/42}. The proton-synchrotron {models} which describe the SED {are} shown in \Cref{fig:hadronic-model} and the corresponding model parameters are reported in \Cref{tab:hadronic-model}.

Hadronic solutions are thus a viable alternative to leptonic ones, and can be achieved with acceptable values for the Doppler factor (equal to $30$) and the jet luminosity (which can be as low as $2 \times 10^{45}$ erg s$^{-1}$, {that is about 0.01 $\times \;L_\text{Edd}$ for a typical supermassive black hole mass of $10^9$ solar masses}).
The emitting region in this case is extremely out of equipartition, being dominated by the magnetic energy density {with {${U_B/U_p \simeq 0.9-120 \times 10^{3}}$}}. The well known degeneracy in the synchrotron radiation spectrum implies that the parameters of the emitting region cannot be constrained, and indeed all studied values of the size of the emitting region from $R_{min} = 10^{14}$ cm, to $R_{max} = 1.46\times  10^{17}$ cm can provide a good solution. The same is true for the values of the magnetic field, which can take any value between $1.2$ and $46.8$ G. The only parameter which takes unusual values is the index of the injected particle population, $\alpha = 1.3$: such a hard injection index is not consistent with standard shock acceleration, although it can be achieved, if particles are accelerated via magnetic reconnection \citep{2014ApJ...783L..21S}. On the other hand, it is important to underline that this value does not come from the SED fitting, but is a direct consequence of the two assumptions on co-acceleration of electrons and protons, and that only synchrotron and adiabatic cooling terms shape the stationary particle distribution. Removing one of these assumptions (or both) can lead to softer particle injection indices still in agreement with the observations.

The hadronic model provides an estimated synchrotron peak located at $\nu_{\text{peak}}^{\text{sync}}  =10^{18}$ Hz ($\simeq 4.1$ keV) and {a gamma-ray peak} located at $\nu_{\text{peak}} = 2-4 \;\; 10^{25}$ Hz ($\simeq0.25$ TeV). {The latter range of values %
 represents the result for the best-fit solutions whose $\chi^2$ is dominated by the \emph{Fermi}-LAT data. 
}

\begin{table}
    \centering
   		\begin{tabular}{@{}l c}
 		\hline
		 & Proton-synchrotron\\
 		\hline
 		 \noalign{\smallskip}
 		$\delta$ & \textit{30} \\
 $R$ [10$^{16}$ cm] & $0.1-14.6$ \\
 $^\star \tau_\textnormal{obs}$ [hours] & $0.3-48.0$ \\
 		\hline
 		 $B$ [G] & $1.2-46.8$   \\
 		$^\star u_B$ [erg cm$^{-3}$] & $0.06-87 $\\
 		\hline
 		$\gamma_{e,\textnormal{min}} $& $200$   \\
 		$\gamma_{e,\textnormal{break}} $& $=\gamma_{e,\textnormal{min}}$     \\
 		$\gamma_{e,\textnormal{max}}\ [10^4]$& $2.5-15.6$\\
 		$\alpha_{e,1}=\alpha_{p,1}$ & $1.3$  \\
 		$\alpha_{e,2}=\alpha_{p,2}$ & $2.3$   \\
 		$K_e$ [10$^{-3}\, $cm$^{-3}$] & $0.015-311$    \\
 		$^\star u_e$ [10$^{-7}\,$erg$\,$cm$^{-3}$] & $0.013-249$\\
 		\hline
 		$\gamma_{p,\textnormal{min}}$& 1  \\
 		$\gamma_{p,\textnormal{break}} [10^9]$&  $=\gamma_{p,\textnormal{max}}$   \\
 		$\gamma_{p,\textnormal{max}} [10^9]$& $2.2-15.7$\\
         $\eta$ [10$^{-5}$]& $0.26-2.6$ \\
 		$^\star u_p$ [10$^{-4}\,$erg cm$^{-3}$] &$0.009-10.7$  \\
 		\hline
    	$^\star U_B/U_p$ [10$^{3}$]& $0.9-120$ \\
 		$^\star L$ [10$^{46}$ erg s$^{-1}$] & $0.2-10.3$  \\
 		\hline
 		\hline
 		\end{tabular}
 	 \newline
\caption{Parameters used for the hadronic model. The luminosity of the emitting region has been calculated as $L=2 \pi R^2c\Gamma_\textnormal{bulk}^2(u_B+u_e+u_p)$, where $\Gamma_\textnormal{bulk}=\delta/2$, and $u_B$, $u_e$, and $u_p$, the energy densities of the magnetic field, the electrons, and the protons, respectively. The quantities flagged with a star ($^\star$) are derived quantities, and not model parameters.}
 		      	 \label{tab:hadronic-model}
 		\end{table}


\section{Conclusions}  
\label{sec:concl}
In this work, we provide the results {of the TeV gamma-ray discovery} of the EHBL  2WHSP~J073326.7+515354. This source has been observed during 2018 with the MAGIC telescopes, which reported a {firm detection} after about 23 hours of observations. Simultaneous data have been collected also in the optical, UV, soft X-ray, and HE gamma-ray bands. This allowed us to build a well-sampled broad-band SED of the source.
{Thanks to the new data for this source, }
we were also able to perform a new estimation of the synchrotron peak and IC peak positions. This leads us to confirm the classification of this source as an EHBL object but showing a softer spectrum {at gamma rays} compared to the ``hard-TeV'' EHBLs like for example 1ES~0229+200.

The broad-band SED has been fitted with four different models: three leptonic and one hadronic. 
The results of leptonic SSC models, whether considering electron acceleration in a spherical plasmoid or in the whole conical expansion of the jet, substantially agree on the extreme spectral parameters needed to fit the SED of this source. The high Doppler factor $\delta$, the low magnetic field $B$ of the emitting region, and the minimum Lorentz factor $\gamma_{min}$, are common resulting parameters. However, for the one-zone leptonic framework an extremely low magnetization is required, {in both models being} very far from equipartition. In order to overcome this problem, two different approaches are used: a two zone leptonic model (spine-layer), and a hadronic scenario. While the one-zone leptonic models result in a ratio between the energy density {of the particles and the magnetic field} ($U_{\rm e}/U_B$) of several orders of magnitude,  the spine-layer model results in a value close to the theoretical expectations. Another interesting point is that equipartition is not reachable with the spine-layer model in other EHBLs like the ``hard-TeV'' blazars (e.g., 1ES 0229+200), and this implies that this object might represent an exception or a transitional case in the EHBL class where the spectral properties are sufficiently extreme but the equipartition regime still holds with respect to hard-TeV blazars (compare with \citealt{Nustar_EHBLs}).

The relatively low luminosity of EHBLs and their modest variability make the application of hadronic modeling successful with reasonable physical parameters. Therefore, in addition to the leptonic models, we presented another model by including a hadronic contribution to the emission mechanism. While the hadronic scenario is able to produce a plausible fit to the MWL SED, the opposite problem for equipartition with respect to one-zone leptonic models is found. The parameter space able to fit the SED results in a ratio $U_B/U_{\rm e}$ far from equipartition, with the jet highly magnetized in this case. 
{For all the models we tested, we ignored the cascade emission by pairs produced in the interaction of TeV photons with the EBL. Emission from these pairs could emerge in the MeV-GeV part of the spectrum if pairs are not isotropized by the intergalactic magnetic field, or if they do not loose energy via other mechanisms. Such an emission, although predicted theoretically, has never been observed so far in any gamma-ray blazar, indicating that, if it exists, it is likely sub-dominant with respect to the emission from the source itself.}
{Finally, considering that the simple one-zone SSC model already provides a good description  of the SED, more complex models (including hadronic component, e.g. a photo-meson model similar to  that discussed in \citealt{2018MNRAS.480..879M}) could only provide second order effects. }

In conclusion, while the four SED modeling scenarios can provide compatible models for the MWL SED of 2WHSP~J073326.7+515354, extreme physical parameters would be required for three of them. The model that better matches with the theoretical predictions is the spine-layer scenario, which provides a reasonable framework to explain the broad-band SED.

The case of 2WHSP~J073326.7+515354, is an important example of the key role that the TeV gamma-ray band plays in the characterization of EHBLs.
New observations of this class of sources by Cherenkov telescopes will allow to increase the number of objects in this population. The forthcoming Cherenkov Telescope Array (CTA) observatory, with its improved sensitivity in this energy band, will be critical in discovering new TeV EHBLs and will help in disclosing the physical phenomena behind their extreme spectral emission.



\section*{Acknowledgements}
%
%
We would like to thank the Instituto de Astrof\'{\i}sica de Canarias for the excellent working conditions at the Observatorio del Roque de los Muchachos in La Palma. The financial support of the German BMBF and MPG, the Italian INFN and INAF, the Swiss National Fund SNF, the ERDF under the Spanish MINECO (FPA2017-87859-P, FPA2017-85668-P, FPA2017-82729-C6-2-R, FPA2017-82729-C6-6-R, FPA2017-82729-C6-5-R, AYA2015-71042-P, AYA2016-76012-C3-1-P, ESP2017-87055-C2-2-P, FPA2017-90566-REDC), the Indian Department of Atomic Energy, the Japanese JSPS and MEXT, the Bulgarian Ministry of Education and Science, National RI Roadmap Project DO1-153/28.08.2018 and the Academy of Finland grant nr. 320045 is gratefully acknowledged. This work was also supported by the Spanish Centro de Excelencia ``Severo Ochoa'' SEV-2016-0588 and SEV-2015-0548, and Unidad de Excelencia ``Mar\'{\i}a de Maeztu'' MDM-2014-0369, by the Croatian Science Foundation (HrZZ) Project IP-2016-06-9782 and the University of Rijeka Project 13.12.1.3.02, by the DFG Collaborative Research Centers SFB823/C4 and SFB876/C3, the Polish National Research Centre grant UMO-2016/22/M/ST9/00382 and by the Brazilian MCTIC, CNPq and FAPERJ.

The \textit{Fermi} LAT Collaboration acknowledges generous ongoing support from a number of agencies and institutes that have supported both the development and the operation of the LAT as well as scientific data analysis. These include the National Aeronautics and Space Administration and the Department of Energy in the United States, the Commissariat \`a l'Energie Atomique and the Centre National de la Recherche Scientifique / Institut National de Physique
Nucl\'eaire et de Physique des Particules in France, the Agenzia Spaziale Italiana and the Istituto Nazionale di Fisica Nucleare in Italy, the Ministry of Education, Culture, Sports, Science and Technology (MEXT), High Energy Accelerator Research Organization (KEK) and Japan Aerospace Exploration Agency (JAXA) in Japan, and the K.~A.~Wallenberg Foundation, the Swedish Research Council and the Swedish National Space Board in Sweden.
 
Additional support for science analysis during the operations phase is gratefully acknowledged from the Istituto Nazionale di Astrofisica in Italy and the Centre National d'\'Etudes Spatiales in France. This work performed in part under DOE Contract DE-AC02-76SF00515.

We acknowledge the use of public data from the \emph{Swift} data archive. This publication makes use of data products from the Wide-field Infrared Survey Explorer, which is a joint project of the University of California, Los Angeles, and the Jet Propulsion Laboratory/California Institute of Technology, funded by the National Aeronautics and Space Administration. {We also acknowledge the use of the Space Science Data Base (SSDC).

J. Becerra Gonz\'alez acknowledges the support of the Viera y Clavijo program funded by ACIISI and ULL. M. Cerruti has received financial support through the Postdoctoral Junior Leader Fellowship Programme from la Caixa Banking Foundation, grant  n. LCF/BQ/LI18/11630012
}



\bibliographystyle{mnras}
\bibliography{bibpaper}



\section*{AFFILIATIONS}

$^{1}$ {Inst. de Astrof\'isica de Canarias, E-38200 La Laguna, and Universidad de La Laguna, Dpto. Astrof\'isica, E-38206 La Laguna, Tenerife, Spain} \\
$^{2}$ {Universit\`a di Udine, and INFN Trieste, I-33100 Udine, Italy} \\
$^{3}$ {National Institute for Astrophysics (INAF), I-00136 Rome, Italy} \\
$^{4}$ {ETH Zurich, CH-8093 Zurich, Switzerland} \\
$^{5}$ {Technische Universit\"at Dortmund, D-44221 Dortmund, Germany} \\
$^{6}$ {Croatian Consortium: University of Rijeka, Department of Physics, 51000 Rijeka; University of Split - FESB, 21000 Split; University of Zagreb - FER, 10000 Zagreb; University of Osijek, 31000 Osijek; Rudjer Boskovic Institute, 10000 Zagreb, Croatia} \\
$^{7}$ {Saha Institute of Nuclear Physics, HBNI, 1/AF Bidhannagar, Salt Lake, Sector-1, Kolkata 700064, India} \\
$^{8}$ {Centro Brasileiro de Pesquisas F\'isicas (CBPF), 22290-180 URCA, Rio de Janeiro (RJ), Brasil} \\
$^{9}$ {IPARCOS Institute and EMFTEL Department, Universidad Complutense de Madrid, E-28040 Madrid, Spain} \\
$^{10}$ {University of \L\'od\'z, Department of Astrophysics, PL-90236 \L\'od\'z, Poland} \\
$^{11}$ {Universit\`a di Siena and INFN Pisa, I-53100 Siena, Italy} \\
$^{12}$ {Deutsches Elektronen-Synchrotron (DESY), D-15738 Zeuthen, Germany} \\
$^{13}$ {Istituto Nazionale Fisica Nucleare (INFN), 00044 Frascati (Roma) Italy} \\
$^{14}$ {Max-Planck-Institut f\"ur Physik, D-80805 M\"unchen, Germany} \\
$^{15}$ {Institut de F\'isica d'Altes Energies (IFAE), The Barcelona Institute of Science and Technology (BIST), E-08193 Bellaterra (Barcelona), Spain} \\
$^{16}$ {Universit\`a di Padova and INFN, I-35131 Padova, Italy} \\
$^{17}$ {Universit\`a di Pisa, and INFN Pisa, I-56126 Pisa, Italy} \\
$^{18}$ {ICRANet-Armenia at NAS RA, 0019 Yerevan, Armenia} \\
$^{19}$ {Centro de Investigaciones Energ\'eticas, Medioambientales y Tecnol\'ogicas, E-28040 Madrid, Spain} \\
$^{20}$ {Universit\"at W\"urzburg, D-97074 W\"urzburg, Germany} \\
$^{21}$ {Finnish MAGIC Consortium: Finnish Centre of Astronomy with ESO (FINCA), University of Turku, FI-20014 Turku, Finland; Astronomy Research Unit, University of Oulu, FI-90014 Oulu, Finland} \\
$^{22}$ {Departament de F\'isica, and CERES-IEEC, Universitat Aut\`onoma de Barcelona, E-08193 Bellaterra, Spain} \\
$^{23}$ {Japanese MAGIC Consortium: ICRR, The University of Tokyo, 277-8582 Chiba, Japan; Department of Physics, Kyoto University, 606-8502 Kyoto, Japan; Tokai University, 259-1292 Kanagawa, Japan; RIKEN, 351-0198 Saitama, Japan} \\
$^{24}$ {Inst. for Nucl. Research and Nucl. Energy, Bulgarian Academy of Sciences, BG-1784 Sofia, Bulgaria} \\
$^{25}$ {Universitat de Barcelona, ICCUB, IEEC-UB, E-08028 Barcelona, Spain} \\
$^{26}$ {also at Port d'Informaci\'o Cient\'ifica (PIC) E-08193 Bellaterra (Barcelona) Spain} \\
$^{27}$ {also at Dipartimento di Fisica, Universit\`a di Trieste, I-34127 Trieste, Italy} \\
$^{28}$ {INAF Istituto di Radioastronomia, via Gobetti 101, 40129 Bologna, Italy} \\

\label{lastpage}


\newpage

\appendix

\section{Observation tables}
In this Appendix, the tables including the MWL observation results are included.  The nightly fluxes obtained in the VHE band as observed by MAGIC telescopes are given in \Cref{table-magic-observations}. The 2-year fluxes obtained in the HE band with the \emph{Fermi}-LAT telescope are reported in  \Cref{table-fermi-lc}. In \Cref{table-swift-observations} and  \Cref{table-swiftUV-observations} we report the information from all the observations of the source with the \emph{Swift}-XRT and \emph{Swift}-UVOT instruments, respectively. The details from the optical observations from KVA telescope are reported in \Cref{tab:kva}.


\begin{table}
    \centering
	\renewcommand{\arraystretch}{1.1} 
		\begin{tabular}{ l c c }
 			\hline
Date	&	Effective time 	&	Flux$_{>\text{200 GeV}}$	\\
MJD	&	s	&$10^{-12}\;\text{ph}\; \text{cm}^{-2}\;\text{s}^{-1}$ 	\\
 			\hline
58141.1 & 2762 & < 9.46		\\
58142.1 & 2584 & < 12.35		\\
58143.1 & 2331 & < 19.45	\\
58144.1 & 2392 & < 16.00	\\	
58164.1 & 2350 & < 8.41	\\		
58165.1 & 2350 & < 3.09	\\		
58167.1 & 2442 & < 8.09	\\		
58169.0 & 3355 & < 12.23\\		
58171.0 & 7938 & 2.97 $\pm$ 1.86 \\
58185.9 & 1363 & < 10.59	\\	
58190.0 & 3126 & < 13.40\\			
58194.0 & 5047 & 3.68  $\pm$  2.51 \\
58194.9 & 4372 & < 10.41 \\	
58195.9 & 2350 & 5.92 $\pm$  3.92 \\
58196.9 & 2350 & < 12.20 \\
58198.9 & 4598 & < 10.64 \\
58199.9 & 5550 & < 7.50 \\
58210.9 & 3496 & < 11.93 \\
58211.9 & 7010 & < 6.75 \\
58213.9 & 4682 & 5.11  $\pm$ 2.77 \\
58226.9 & 2342 & < 16.05 \\
	 			\hline
		\end{tabular}
	 \caption{Flux (points and 95\% C.L. upper limits) and effective observing time of the source 2WHSP~J073326.7+515354 registered by the MAGIC telescopes.}
	\label{table-magic-observations}
\end{table}

\begin{table*}
	\begin{center}
	\renewcommand{\arraystretch}{1.1} 
		\begin{tabular}{cccccc}
 			\hline
$MJD_{start}$ & $MJD_{stop}$	&	Flux  & $\Gamma$ & TS \\
	& &	 $10^{-10}\;\text{ph cm}^{-2}\;\text{s}^{-1}$&  	& \\
 			\hline
54682.7& 55412.7    &    2.6 $\pm$ 1.1 & 1.4 $\pm$ 0.2 & 45.6\\
55412.7& 56142.7&        3.2 $\pm$ 1.4 & 1.6 $\pm$ 0.2 & 25.1\\ 
56142.7& 56872.7    &    6.8 $\pm$ 2.0 & 1.8 $\pm$ 0.2 & 58.8\\
56872.7& 57602.7&        1.9 (U.L.)     &           1.7 & 0.4\\
57602.7& 58332.7&        1.6 $\pm$ 1.1 & 1.5 $\pm$ 0.4 & 15.9\\
\hline
		\end{tabular}
	\end{center}
	 \caption{\textit{Fermi}-LAT light-curve generated for 2-year time bins {within the 0.5-300 GeV energy band}. In case of non significant detection (TS<4), a 95\% C.L. flux upper-limit was estimated assuming the spectral index reported in the 4FGL catalog.}
	\label{table-fermi-lc}
\end{table*}

\begin{table*}
\renewcommand{\arraystretch}{1.37} 
\centering
\begin{tabular}{lcccccccccccc}
\hline
 Date & MJD	&	 Exposure &  F$_{2-10keV}$     &  F$_{0.3-10keV}$	   & $\Gamma_{\text{X}}$	& $\chi ^2$/d.o.f. & Obs. ID  \\
 &		&	s	   & $10^{-11}$\;erg\;cm$^{-2}$\;s$^{-1}$	&  $10^{-11}$\;erg\;cm$^{-2}$\;s$^{-1}$  &			&		&  \\
\hline
2009-12-30&55195.8	& 1061 & $1.36\pm0.16$ & $2.12\pm0.17$ & $1.71\pm0.08$ & 15.59/17   & 00038675001 \\
2011-02-20&55612.2	& 9872 & $0.84\pm0.03$ & $1.34\pm0.04$ & $1.76\pm0.03$ & 117.16/112 & 00045364001 \\
2011-02-24&55616.4	& 492  & $0.57\pm0.18$ & $1.13\pm0.19$ & $2.07\pm0.21$ & 0.38/3	    & 00045364002 \\
2014-01-11&56668.7	& 1096 & $1.90\pm0.16$ & $2.99\pm0.16$ & $1.73\pm0.06$ & 20.17/25   & 00048299002 \\
2014-01-12&56669.7	& 994  & $1.91\pm0.17$ & $2.89\pm0.19$ & $1.67\pm0.07$ & 17.44/21   & 00048299003 \\
\hline
2018-01-26&58144.1	& 1326 & $1.32\pm0.11$ & $1.95\pm0.10$ & $1.62\pm0.06$ & 27.66/25   & 00010541001 \\
2018-02-07&58156.0	& 1364 & $1.03\pm0.09$ & $1.72\pm0.09$ & $1.82\pm0.06$ & 22.24/26   & 00010541002 \\
2018-02-15&58164.0	& 1059 & $1.62\pm0.20$ & $2.27\pm0.19$ & $1.51\pm0.09$ & 20.33/16   & 00010541003 \\
2018-02-22&58171.9	& 1004 & $1.80\pm0.19$ & $2.58\pm0.20$ & $1.55\pm0.07$ & 19.94/18   & 00010541004 \\
2018-03-06&58183.9	& 1441 & $1.14\pm0.10$ & $1.79\pm0.11$ & $1.72\pm0.07$ & 26.28/25   & 00010541006 \\
2018-03-12&58190.0	& 1419 & $1.40\pm0.12$ & $2.08\pm0.13$ & $1.62\pm0.06$ & 22.83/29   & 00010541007 \\
2018-04-08&58216.4	& 1094 & $1.42\pm0.12$ & $2.18\pm0.14$ & $1.68\pm0.07$ & 29.56/23   & 00010541009 \\
2018-04-19&58227.9	& 1136 & $1.09\pm0.09$ & $1.84\pm0.11$ & $1.84\pm0.07$ & 38.97/24   & 00010541010 \\
\hline
\end{tabular}
	 \caption{Results of {\it Swift}-XRT data analysis for the observations of 2WHSP~J073326.7+515354. We report for each observation the MJD, the exposure obtained by XRT, and the two integral fluxes between 2-10 keV and 0.3-10 keV. Every individual spectrum can be well fitted by a power-law function with spectral index $\Gamma_{\text{X}}$ and good reduced $\chi^2$. In addition to the simultaneous observations with MAGIC, the results from historical observations are also included for comparison purposes.}
	\label{table-swift-observations}
\end{table*}

\begin{table*}
\renewcommand{\arraystretch}{1.37} 
\centering
\begin{tabular}{lcccccccccccc}
\hline
Band & Date & MJD	&	 Flux  & Compatibility \\
 &  & 	&$10^{-12}$\;erg\;cm$^{-2}$\;s$^{-1}$	   &with average flux \\
\hline
U	&	2011-02-20	&	55612	&	1.79	$\pm$	0.10	&	0.4		\\
	&	2011-02-24	&	55616	&	1.76	$\pm$	0.11	&	0.1		\\
	&	2014-01-11	&	56668	&	2.38	$\pm$	0.10	&	5.8		\\
	&	2018-01-26	&	58144	&	1.74	$\pm$	0.08	&	0.1		\\
	&	2018-02-07	&	58156	&	1.82	$\pm$	0.09	&	0.7		\\
	&	2018-02-15	&	58164	&	1.63	$\pm$	0.10	&	1.1		\\
	&	2018-04-08	&	58216	&	1.88	$\pm$	0.11	&	1.1		\\
												
W1	&	2009-12-30	&	55195	&	1.40	$\pm$	0.11	&	0.6		\\
	&	2018-02-22	&	58171	&	1.26	$\pm$	0.10	&	0.0		\\
	&	2018-03-06	&	58183	&	1.21	$\pm$	0.11	&	0.2		\\
	&	2018-04-19	&	58227	&	1.29	$\pm$	0.11	&	0.1		\\
												
W2	&	2011-02-20	&	55612	&	1.15	$\pm$	0.10	&	0.3		\\
	&	2014-01-12	&	56669	&	1.83	$\pm$	0.11	&	4.6		\\
	&	2018-03-12	&	58189	&	1.18	$\pm$	0.08	&			\\
\hline
\end{tabular}
	 \caption{Results of {\it Swift}-UV data analysis for the observations of 2WHSP~J073326.7+515354. We report the energy band of the UVOT instrument, the date and MJD of the observation, the integral flux, and the compatibility between each value and the average flux for only 2018 data simultaneous to MAGIC observations. The compatibility has been computed as $\lambda=\frac{|A-B|}{\sqrt[]{\sigma_A^2+\sigma_B^2}}$, { for a given two fluxes with values A and B and their respective uncertainties ($\sigma_A$ and $\sigma_B$).}
	 where for example $A$ is a flux value and $B$ is the average flux reported in the text. }
	\label{table-swiftUV-observations}
\end{table*}

\begin{table}
    \centering
    	\renewcommand{\arraystretch}{1.1} 
    \begin{tabular}{lc}
    \hline
        MJD & Flux \\
                 & $10^{-4}$ Jy \\
        \hline
58210.500	&	4.01	$\pm$	0.38	\\
58211.462	&	3.87	$\pm$	0.39	\\
58212.467	&	3.89	$\pm$	0.39	\\
58214.479	&	3.94	$\pm$	0.41	\\
58215.455	&	3.61	$\pm$	0.60	\\
58218.508	&	3.85	$\pm$	0.37	\\
58219.465	&	3.48	$\pm$	0.40	\\
58222.487	&	4.10	$\pm$	0.40	\\
58226.473	&	3.55	$\pm$	0.39	\\
58232.497	&	3.84	$\pm$	0.41	\\
58233.469	&	2.98	$\pm$	0.44	\\
58241.482	&	4.00	$\pm$	0.39	\\
58403.851	&	4.07	$\pm$	0.38	\\
58423.835	&	3.98	$\pm$	0.39	\\
58471.822	&	4.41	$\pm$	0.38	\\
58478.822	&	4.09	$\pm$	0.38	\\
58481.804	&	4.00	$\pm$	0.38	\\
58482.743	&	3.94	$\pm$	0.38	\\
58488.802	&	3.92	$\pm$	0.37	\\
\hline
    \end{tabular}
    \caption{Optical R-band flux of 2WHSP~J073326.7+515354 as measured by the KVA telescope. The data have already been corrected by host galaxy subtraction.}
    \label{tab:kva}
\end{table}
\qquad\\

\section{Estimation of the broad-band SED peak frequency}

\begin{figure}[h]
\centering
\subfloat[][Synchrotron peak of 2WHSP~J073326.7+515354. The log-parabolic model fits the simultaneous \emph{Swift}-XRT data  (violet  dots) and the archival  \emph{Swift}-BAT 105-months catalogue \citep{BATcatalog105} data  (dark red dots) .]{\includegraphics[width=0.8\columnwidth]{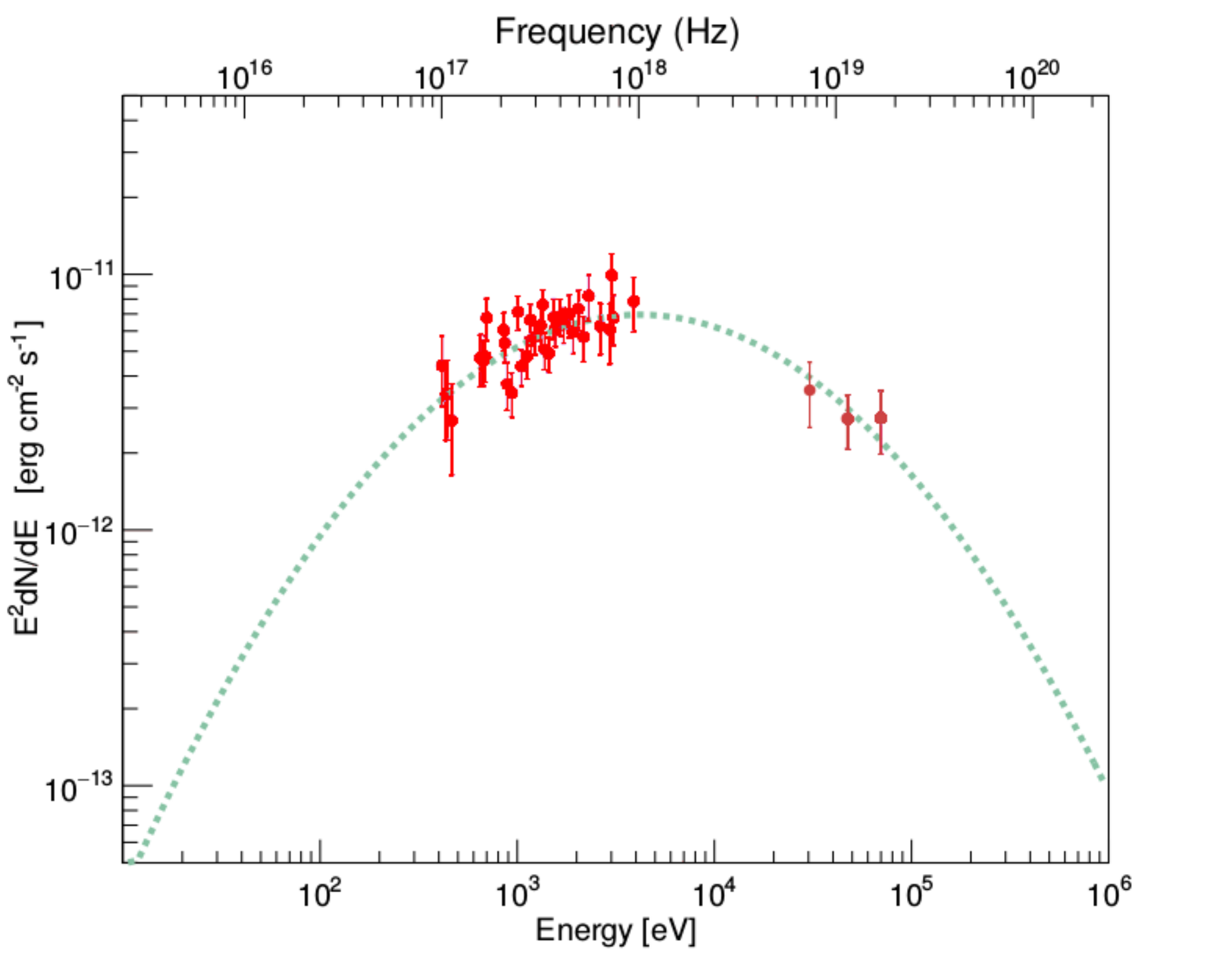}\label{fig:synchrotronpeak}
}\\
\subfloat[][High energy SED hump of 2WHSP~J073326.7+515354. The MAGIC spectral points (blue dots) are EBL-corrected using the model from \citet{Dominguez:2010bv}. {A power-law with an exponential cut-off is used to fit the \emph{Fermi}-LAT data (red circles) and the MAGIC data (blue circles)}]{\includegraphics[width=0.8\columnwidth]{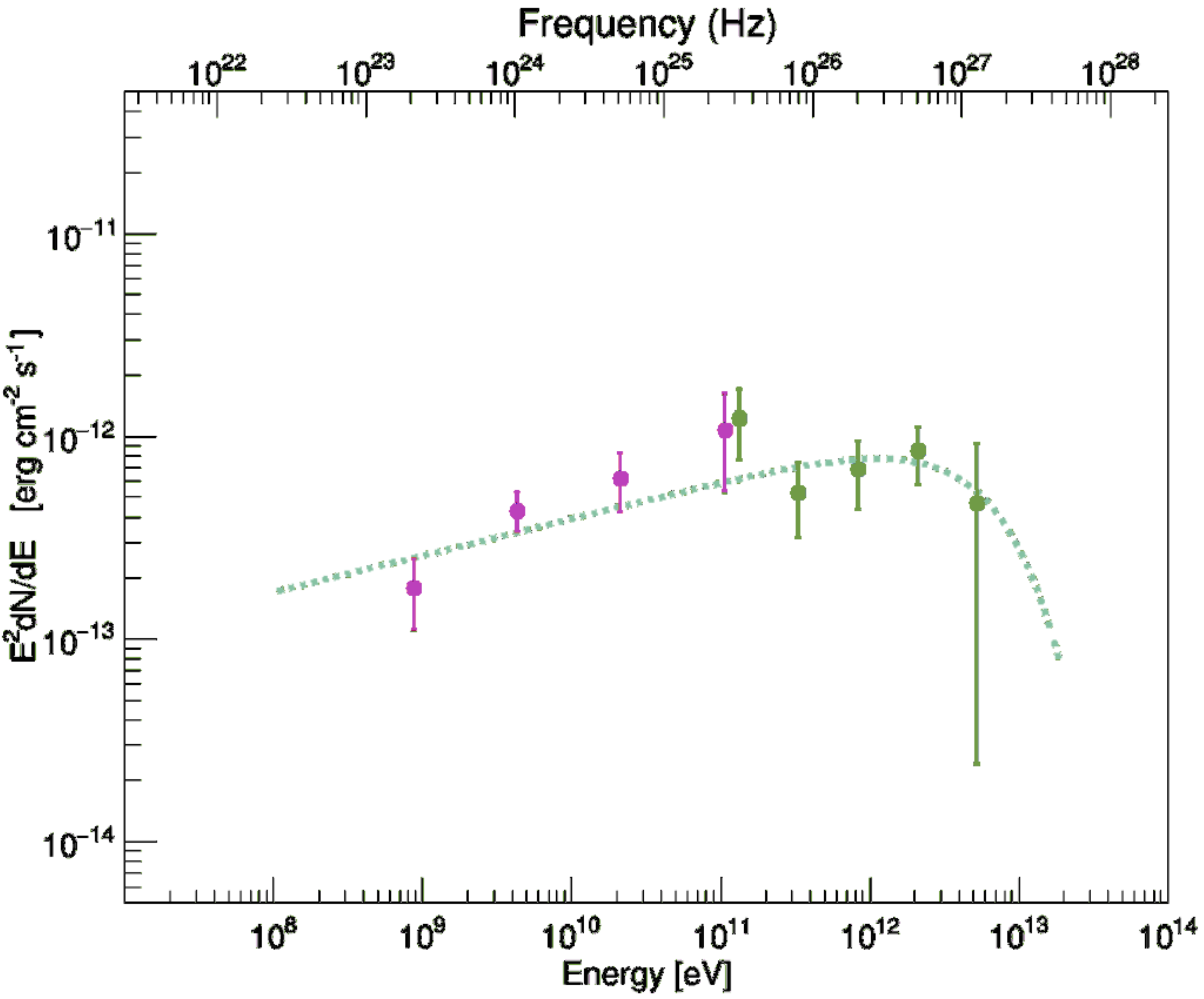}\label{fig:ICpeak}
}
\caption{Synchrotron (a) and (b) IC hump of  2WHSP~J073326.7+515354.}
\label{fig:synchroICpeak}
\end{figure}

The frequencies at which the peaks of the MWL SED are located are crucial for the classification of the target. For the case of 2WHSP~J073326.7+515354, the low energy peak is characterized by means of the {\it Swift}-XRT and {\it Swift}-BAT spectra. The high energy peak instead is characterized by the gamma-ray observations from {\it Fermi}-LAT and MAGIC. The SEDs can be found in \Cref{fig:synchroICpeak}.

\end{document}